\documentclass[10pt,aps,prb,twocolumn,notitlepage,showpacs,superscriptaddress]{revtex4-1}
\usepackage{amsthm,amsmath,amsfonts,amssymb,verbatim,xcolor}
\usepackage[charter]{mathdesign}
\usepackage{graphicx}
\usepackage{subfigure}
\usepackage{bm}
\usepackage{epsfig,slashed}
\usepackage[T1]{fontenc}
\usepackage[colorlinks=true,citecolor=blue,linkcolor=blue,urlcolor=blue]{hyperref}
\newcommand{\tr}{\mathop{\mathrm{Tr}}}

\renewcommand{\b}[1]{{\boldsymbol{#1}}}

\newcommand{\sgn}{\mathop{\mathrm{sgn}}}

\newcommand\s{\sigma}

\newcommand{\Gv}{\b{G}}
\newcommand{\Sigmav}{\b{\Sigma}}
\newcommand{\Vv}{\b{V}}

\DeclareGraphicsExtensions{.png}

\begin{document}

\title{Quantum cluster approach to the spinful Haldane-Hubbard model}

\author{Jingxiang Wu}
\affiliation{Department of Physics, University of Alberta, Edmonton, Alberta T6G 2E1, Canada}

\author{Jean Paul Latyr Faye}
\affiliation{D\'{e}partement de physique, Universit\'{e} de Sherbrooke, Sherbrooke, Qu\'{e}bec J1K 2R1, Canada}

\author{David S\'{e}n\'{e}chal}
\affiliation{D\'{e}partement de physique, Universit\'{e} de Sherbrooke, Sherbrooke, Qu\'{e}bec J1K 2R1, Canada}

\author{Joseph Maciejko}
\email[electronic address: ]{maciejko@ualberta.ca}
\affiliation{Department of Physics, University of Alberta, Edmonton, Alberta T6G 2E1, Canada}
\affiliation{Theoretical Physics Institute, University of Alberta, Edmonton, Alberta T6G 2E1, Canada}
\affiliation{Canadian Institute for Advanced Research, Toronto, Ontario M5G 1Z8, Canada}

\date\today

\begin{abstract}
We study the spinful fermionic Haldane-Hubbard model at half filling using a combination of quantum cluster methods: cluster perturbation theory (CPT), the variational cluster approximation (VCA), and cluster dynamical mean-field theory (CDMFT). We explore possible zero-temperature phases of the model as a function of on-site repulsive interaction strength and next-nearest-neighbor hopping amplitude and phase. Our approach allows us to access the regime of intermediate interaction strength, where charge fluctuations are significant and effective spin model descriptions may not be justified. Our approach also improves upon mean-field solutions of the Haldane-Hubbard model by retaining local quantum fluctuations and treating them nonperturbatively. We find a correlated topological Chern insulator for weak interactions and a topologically trivial N\'eel antiferromagnetic insulator for strong interactions. For intermediate interactions, we find that topologically nontrivial N\'eel antiferromagnetic insulating phases and/or a topologically nontrivial nonmagnetic insulating phase may be stabilized.
\end{abstract}


\maketitle

\section{Introduction}

A recent paper\cite{jotzu2014} has reported the realization of Haldane's model of the quantum anomalous Hall effect,\cite{haldane1988} or quantum Hall effect without Landau levels, in a system of ultracold fermionic ${}^{40}$K atoms loaded into a honeycomb optical lattice. While other recent realizations of the quantum anomalous Hall effect in Cr-doped (Bi,Sb)$_2$Te$_3$ topological insulator thin films\cite{chang2013,bestwick2015,chang2015} are equally impressive, Jotzu \emph{et al.}'s optical-lattice setup\cite{jotzu2014} opens up unique possibilities for the simulation of quantum models of \emph{correlated} particles with topological bandstructures, i.e., topological versions of the Bose-Hubbard\cite{greiner2002} or Fermi-Hubbard\cite{kohl2005} models. Although topological band insulators are inherently stable against sufficiently weak symmetry-preserving interactions, their fate in the presence of strong interactions remains a largely unsolved but actively investigated problem.\cite{InteractingTI}

Motivated by Ref.~\onlinecite{jotzu2014}, in this paper we aim to determine the ground-state phase diagram of the half-filled spinful Haldane-Hubbard (HH) model.\cite{he2011,he2011b,he2012,he2012b,maciejko2013,zhu2014,hickey2015,zheng2015,prychynenko2014,hickey2015b,arun2015} This model can be seen as a hybrid of the standard Haldane and Hubbard models, in which spin-1/2 fermions hop on a two-dimensional (2D) honeycomb lattice according to Haldane's original tight-binding model, but also repel each other when on the same site with energy cost $U$ (Fig.~\ref{fig:lattice}). Unlike its time-reversal invariant counterpart the Kane-Mele-Hubbard model, which has been studied successfully by quantum Monte Carlo (QMC) methods in recent years,\cite{hohenadler2011,zheng2011,hohenadler2012,hohenadler2012b,assaad2013,hung2013,lang2013,meng2013,lai2014,hung2014,bercx2014,lai2014b,ma2014,wu2015} the HH model breaks time-reversal symmetry, which leads to the notorious fermion sign problem and precludes the use of QMC methods. Previous studies of the HH model have thus investigated two limiting cases. The first is the large-$U$ limit, in which the HH model at half filling is mapped to an effective $\mathrm{SU}(2)$-invariant spin model involving only the spin degree of freedom of the original fermions. The resulting spin model has been studied using classical variational approaches\cite{hickey2015} as well as exact diagonalization (ED) on small clusters.\cite{hickey2015b} Although the large-$U$ limit is expected to give a good description of the physics deep in the Mott insulating phase where charge degrees of freedom are frozen, it cannot describe the Mott/symmetry-breaking transitions out of the weakly interacting Chern insulating phase. It may even fail in the weak Mott regime, i.e., on the Mott side but close enough to the transition, where charge fluctuations are pronounced and the usual procedure of keeping only a few terms in the $t/U$ expansion ($t$ is a characteristic hopping amplitude) may not be justified. Another line of attack has been to treat the full fermionic HH model with both charge and spin degrees of freedom, but to neglect quantum fluctuations entirely and use a mean-field approach so that the problem remains tractable. This approach has been used either in the context of conventional Hartree-Fock theory\cite{he2011,he2011b,he2012,he2012b,zhu2014,zheng2015,prychynenko2014} or slave-particle mean-field theory.\cite{he2011,he2011b,maciejko2013,hickey2015,prychynenko2014} Here one has the advantage over the large-$U$ limit of being able to describe transitions out of the Chern insulating phase and the intermediate $U$ regime, but at the expense of neglecting quantum fluctuations which can quantitatively and qualitatively influence the ground-state phase diagram of the model.

Here we use a combination of quantum cluster methods\cite{senechal2008} --- cluster perturbation theory\cite{gros1993,senechal2000} (CPT), the variational cluster approximation\cite{potthoff2003} (VCA), and cluster dynamical mean-field theory\cite{Lichtenstein:2000vn,Kotliar:2001} (CDMFT) --- to study the half-filled spinful HH model. Quantum cluster methods are one of the few methods that allow us to study the full interacting fermionic problem for all values of $U$, while taking quantum fluctuations into account nonperturbatively. Furthermore, these methods are formulated directly in the thermodynamic limit. While they do not present an exact solution to the problem, quantum cluster methods capture the full dynamical (i.e., frequency-dependent) effect of short-range correlations and thus constitute a significant improvement over mean-field approaches. In these methods, one views the lattice of the original problem as a superlattice of small clusters connected by hopping. The size of the clusters is chosen such that the problem of decoupled clusters can be solved by numerical exact diagonalization (ED). An approximate solution to the original problem of coupled clusters is then obtained by treating hopping between clusters in perturbation theory to infinite order, in the spirit of strong-coupling perturbation theory.\cite{pairault1998} Tendencies towards symmetry-breaking long-range order can then be studied by means of a dynamical variational principle for correlated systems, Potthoff's self-energy-functional theory.\cite{potthoff2003b} Quantum cluster methods have been used successfully in the study of correlated topological phases of matter, including correlated Chern insulators,\cite{hassan2013,nguyen2013} quantum spin Hall insulators,\cite{yu2011,yoshida2012,tada2012,cocks2012,budich2012,wu2012,yoshida2013,miyakoshi2013,budich2013,nourafkan2014,laubach2014,chen2015,grandi2015} topological Kondo insulators,\cite{werner2013} and Weyl semimetals.\cite{witczak-krempa2014}

Our main findings can be summarized as follows: (1) the ground-state phase diagram contains a topologically trivial N\'eel antiferromagnetic (AF) insulator at large $U$ and a correlated topological Chern insulator (CI) at small $U$, in agreement with previous Hartree-Fock findings; (2) in both VCA and CDMFT, topologically nontrivial N\'eel AF phases appear at intermediate $U$; (3) in VCA, a topologically nontrivial nonmagnetic insulator (NMI) appears for intermediate $U$, sandwiched between the CI and AF phases, while in CDMFT the NMI phase is preempted by the onset of AF order as the interaction strength $U$ increases. Topologically nontrivial phases are characterized by a nonzero value of the generalized Chern number computed from the one-particle Green's function.

The rest of the paper is organized as follows. In Sec.~\ref{sec:model} we introduce the HH model; in Sec.~\ref{sec:vca} we give a brief introduction to quantum cluster methods; in Sec.~\ref{sec:results} and \ref{sec:cdmft} we present our VCA and CDMFT results, respectively; and in Sec.~\ref{sec:concl} we discuss these results as well as possible avenues for future work.

\section{Spinful Haldane-Hubbard model}
\label{sec:model}

\begin{figure}[t]
\includegraphics[scale=0.9]{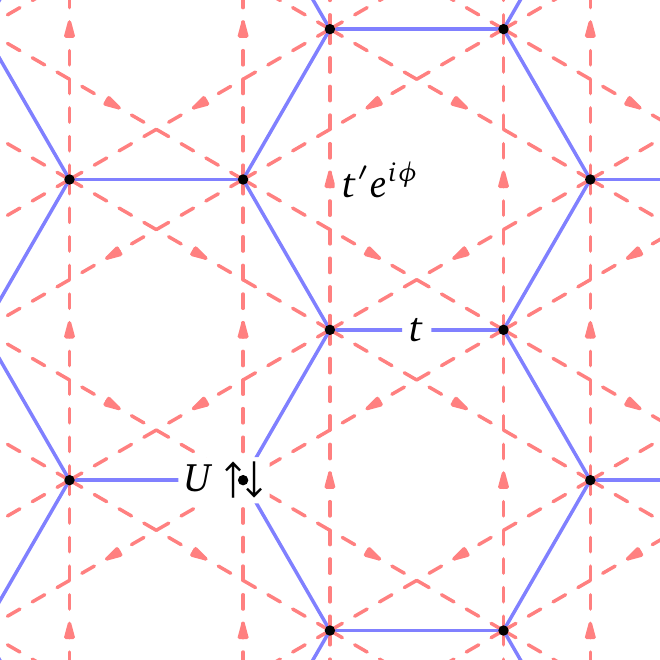}
\caption{(Color online) Schematic depiction of the spinful Haldane-Hubbard model on the honeycomb lattice: the real nearest-neighbor hopping $t$ and complex next-nearest-neighbor hopping $t'e^{\pm i\phi}$ give rise to the band structure of a Chern insulator, while the on-site repulsion $U>0$ introduces electronic correlations.}\label{fig:lattice}
\end{figure}

The half-filled spinful HH model\cite{he2011,he2011b,he2012,he2012b,maciejko2013,zhu2014,hickey2015,zheng2015,prychynenko2014} is defined by the Hamiltonian (Fig.~\ref{fig:lattice})
\begin{align}\label{H}
H(t'/t,U/t,\phi)=&-t\sum_{\langle ij\rangle\sigma}c_{i\sigma}^\dag c_{j\sigma}-t'\sum_{\langle\langle ij\rangle\rangle\sigma}e^{i\nu_{ij}\phi}c_{i\sigma}^\dag c_{j\sigma}\nonumber\\
&+U\sum_i n_{i\uparrow}n_{i\downarrow}-\mu\sum_{i\sigma}c_{i\sigma}^\dag c_{i\sigma},
\end{align}
where $c_{i\sigma}^\dag$ ($c_{i\sigma}$) creates (annihilates) an electron of spin $\sigma=\uparrow,\downarrow$ on site $i$ of the two-dimensional (2D) honeycomb lattice, $t$ is the nearest-neighbor hopping amplitude, $t'e^{\nu_{ij}\phi}$ is the next-nearest-neighbor hopping amplitude with $\nu_{ij}=+1$ ($-1$) for clockwise (counter-clockwise) hopping and we adopt the convention that $-\pi<\phi\leq\pi$, $U>0$ is the on-site repulsion energy, and the chemical potential $\mu$ is chosen to maintain the system at half filling. Under time-reversal symmetry ($\mathcal{T}$), the Hamiltonian transforms as
\begin{align}\label{TRS}
\mathcal{T}H(t'/t,U/t,\phi)\mathcal{T}^{-1}=H(t'/t,U/t,-\phi),
\end{align}
hence any $\phi\neq 0,\pi$ breaks $\mathcal{T}$ explicitly. In the noninteracting limit $U=0$, Eq.~(\ref{H}) reduces to two decoupled copies of Haldane's original model\cite{haldane1988} for the quantum Hall effect without Landau levels, and describes a gapped Chern insulator (CI) with total Chern number $C=2\sgn\phi$ and quantized Hall conductivity $\sigma_{xy}=Ce^2/h$ for $\phi\neq 0,\pi$. The system becomes gapless for $\phi=0,\pi$, where time-reversal symmetry is restored. At half filling, particle-hole symmetry ($\mathcal{C}$) implies
\begin{align}
\mathcal{C}H(t'/t,U/t,\phi)\mathcal{C}^{-1}=H(t'/t,U/t,\pi-\phi),
\end{align}
which, combined with Eq.~(\ref{TRS}) and the fact that a change of sign of $t'$ is equivalent to a shift of $\phi$ by $\pm\pi$, implies
\begin{align}
\mathcal{C}\mathcal{T}H(t'/t,U/t,\phi)(\mathcal{C}\mathcal{T})^{-1}=H(-t'/t,U/t,\phi).
\end{align}
Thus the phase diagram of the interacting model is symmetric about $t'=0$ and $\phi=\pi/2$, and it is sufficient to study the problem for $t'\geq 0$ and $0\leq\phi\leq\pi/2$. Henceforth all energies will be measured in units of $t$, and we set $t=1$.

Previous mean-field studies\cite{he2011,he2011b,he2012,he2012b,maciejko2013,zhu2014,hickey2015,zheng2015,prychynenko2014,arun2015} all agree on the fact that for $\phi\neq 0,\pi$, the CI is stable against the Hubbard interaction for $U$ less than some critical $U_c(\phi)$, where the detailed form of $U_c(\phi)$ depends on the approach being used. One expects the CI to be perturbatively stable against interactions because it is a gapped state. However, different approaches lead to different conclusions regarding what phases occur for $U>U_c(\phi)$, and in what order. Most conventional mean-field studies predict the occurrence of magnetically ordered phases at large enough $U$,\cite{zhu2014,hickey2015,zheng2015,prychynenko2014,arun2015} while slave-particle studies predict additional nonmagnetic, topologically ordered phases at intermediate $U$, such as the chiral spin liquid (CSL)\cite{he2011,hickey2015,kalmeyer1987,*kalmeyer1989,
wen1989,schroeter2007,thomale2009} and the correlated Chern insulator (CI*).\cite{maciejko2013,prychynenko2014} The possibility of a CSL ground state in models of CIs augmented by an on-site Hubbard repulsion can also be inferred from Gutzwiller-projection studies of such models\cite{zhang2011} and ED studies of effective spin Hamiltonians in the large-$U$ limit.\cite{nielsen2013,hickey2015b}

We will focus on the possibility of AF order in the spinful HH model, as well as possible non-symmetry-breaking Mott transitions. By contrast with studies of effective spin Hamiltonians valid deep in the Mott phase $U\gg 1$ where charge fluctuations are completely frozen,\cite{nielsen2013,hickey2015,hickey2015b} here charge fluctuations remain included and we are able to access Mott transitions and the weak Mott regime $U\sim 1$. Given the bipartite nature of the honeycomb lattice and the $\mathrm{SU}(2)$ spin symmetry of the model, AF order is the simplest and most likely type of order one can consider. Motivated in particular by the Hartree-Fock study of Ref.~\onlinecite{zheng2015} which finds collinear (N\'{e}el) AF order in this model for sufficiently large $U$ and $t'\lesssim 0.35$, we will focus on the range $0<t'<0.35$ and likewise evaluate the likelihood of N\'{e}el AF order as a function of $U>0$ and $0<\phi<\pi/2$. N\'{e}el AF order is also a natural choice because the $t'=0$ limit of the model corresponds to the nearest-neighbor Hubbard model on the honeycomb lattice, which has been convincingly shown to exhibit N\'{e}el AF order for $U>3.869$ via large-scale, sign-problem-free QMC simulations.\cite{sorella2012}

\section{Quantum cluster methods}
\label{sec:vca}

\begin{figure}[t]
\includegraphics[scale = 0.9]{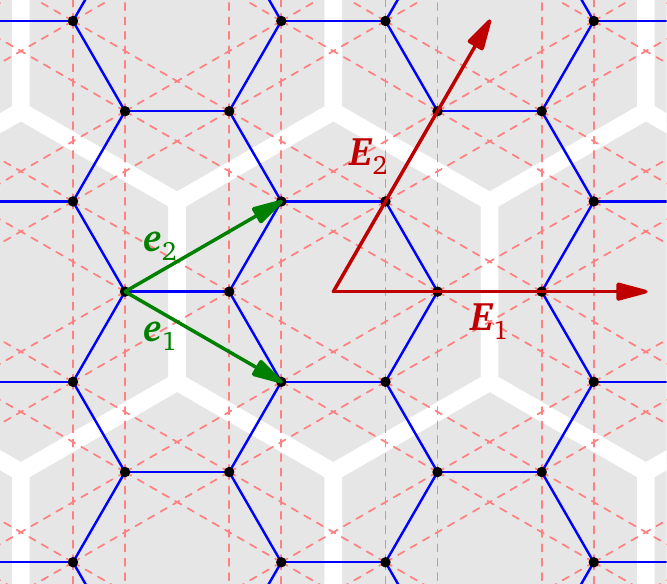}
\caption{(Color online) Six-site cluster used in the VCA calculations (shaded area), with Bravais lattice vectors $\b{e}_1,\b{e}_2$ (green arrows) and superlattice vectors $\b{E}_1,\b{E}_2$ (red arrows).}
  \label{fig:cluster}
\end{figure}

Quantum cluster methods are based on Potthoff's variational principle for strongly correlated systems.\cite{potthoff2003,potthoff2003b} Given a Hamiltonian $H=H_0(\b{t})+H_1(U)$ that is the sum of a noninteracting term $H_0(\b{t})$ with one-body Hamiltonian matrix $\b{t}$, and a local interaction term $H_1(U)$ with interaction strength $U$, one defines a functional $\Omega_\b{t}[\Sigmav]$ of the self-energy $\Sigmav$ as
\begin{align}\label{Potthoff1}
\Omega_\b{t}[\Sigmav]=\tr\ln\left(-\left(\Gv_0^{-1}-\Sigmav\right)^{-1}\right)
+F[\Sigmav],
\end{align}
where the trace and logarithm are to be understood in the functional sense, $\Gv_0=(\omega+\mu-\b{t})^{-1}$ is the one-particle Green's function of the noninteracting system, and $F[\Sigmav]=\Phi[\Gv[\Sigmav]]-\tr(\Sigmav \Gv[\Sigmav])$ is the Legendre transform of the Luttinger-Ward functional $\Phi[\Gv]$,\cite{luttinger1960} $\Gv$ being regarded as a functional of $\Sigmav$. Potthoff's principle states that $\Omega_\b{t}[\Sigmav]$ is stationary at the exact (physical) self-energy, and its value at the stationary point coincides with the exact thermodynamic grand potential $\Omega$ of the system.

Because the exact form of $F[\Sigmav]$ is not known in general, one cannot directly use Eq.~(\ref{Potthoff1}) for variational calculations. However, one can take advantage of the fact that the functional form of $F[\Sigmav]$ depends only on the interaction term $H_1(U)$ and not on the one-body term $H_0(\b{t})$. $F[\Sigmav]$ inherits this property from $\Phi[\Gv]$, whose diagrammatic representation contains only skeleton diagrams with fully dressed Green's functions $\Gv$ and interaction vertices $U$, but no explicit dependence on $\b{t}$. One thus defines a reference Hamiltonian $H'=H_0(\b{t}')+H_1(U)$ that differs from $H$ in its one-body Hamiltonian matrix $\b{t}'$ only, but which is more easily solved. In the context of quantum cluster methods, one chooses $\b{t}'$ by severing bonds in $H_0$ such $H'$ describes fully interacting but decoupled clusters. One can then use ED to compute the fully interacting one-particle Green's function $\Gv'$ for the reference Hamiltonian, from which the exact self-energy $\Sigmav(\b{t}')$ and grand potential $\Omega'$ for $H'$ can be determined. Applying Eq.~(\ref{Potthoff1}) to the reference problem, we obtain
\begin{align}\label{Potthoff2}
\Omega_{\b{t}'}[\Sigmav(\b{t}')]=\Omega'=\tr\ln\left(-\left(\Gv_0^{\prime -1}-\Sigmav(\b{t}')\right)^{-1}\right)+F[\Sigmav(\b{t}')],
\end{align}
where $\Gv_0'=(\omega+\mu-\b{t}')^{-1}$ is the noninteracting Green's function for the reference Hamiltonian. We used the fact that the functional form of $F[\Sigmav]$ is the same regardless of the one-body term, and the fact that $\Sigmav(\b{t}')$ is a stationary point of $\Omega_{\b{t}'}[\Sigmav]$ since it is the exact self-energy for $H'$. Equation~(\ref{Potthoff2}) can then be used to give an explicit expression for $F[\Sigmav]$:
\begin{align}
F[\Sigmav(\b{t}')]=\Omega'-\tr\ln\left(-\Gv'\right),
\end{align}
using $\Gv^{\prime -1}=\Gv_0^{\prime -1}-\Sigmav(\b{t}')$.

We now assume that the exact self-energy of the original Hamiltonian $H$ can be represented as the self-energy $\Sigmav(\b{t}')$ of the reference Hamiltonian $H'$ for a suitable choice of $\b{t}'$. In other words, we search for a stationary point of $\Omega_\b{t}[\Sigmav]$ on the set of self-energies of this form. Using Eq.~(\ref{Potthoff1}) and (\ref{Potthoff2}), the functional to be extremized is
\begin{align}\label{Potthoff3}
\Omega_\b{t}[\Sigmav(\b{t}')]&= \Omega'+\tr\ln\left(-\left(\Gv_0^{-1}-\Sigmav(\b{t}')
\right)^{-1}\right)\nonumber\\
&\hspace{5mm}-\tr\ln(-\Gv').
\end{align}
In practice, one extremizes the functional (\ref{Potthoff3}) with respect to the one-body Hamiltonian matrix $\b{t}'$ of the decoupled clusters.

So far the discussion has been exact, assuming the exact self-energy is $\b{t}'$-representable as explained earlier. In CPT,\cite{gros1993,senechal2000} one approximates the exact Green's function $\Gv$ of the original Hamiltonian $H$ as
\begin{align}\label{GCPT}
\Gv^{-1}=\Gv_0^{-1}-\Sigmav(\b{t}')=\Gv^{\prime -1}-\Vv,
\end{align}
where $\Vv=\b{t}-\b{t}'$ corresponds to inter-cluster hopping terms that were severed in the reference Hamiltonian. The cluster Green's function $\Gv'$ is thus viewed as the unperturbed Green's function, and $\Vv$ is treated as a perturbation (albeit to infinite order). Using Eq.~(\ref{GCPT}), the Potthoff functional (\ref{Potthoff3}) can be written as
\begin{align}\label{Potthoff4}
\Omega_\b{t}[\Sigmav(\b{t}')]=\Omega'-\tr\ln\left(1-\Vv\Gv'\right).
\end{align}
In VCA, one searches for stationary points of the functional (\ref{Potthoff4}), i.e., solutions of the Euler equation $\partial\Omega_\b{t}[\Sigmav(\b{t}')]/\partial\b{t}'=0$. This is achieved in practice by using the cluster one-body terms $\b{t}'$ as variational parameters. In particular, one can search for spontaneously broken symmetries by including in $\b{t}'$ symmetry-breaking terms, i.e., Weiss fields. By contrast with conventional mean-field theory however, here the full dynamical effect of correlations is taken into account via the frequency dependence of the cluster Green's function $\Gv'$ in Eq.~(\ref{Potthoff4}).

We choose the reference Hamiltonian $H'$ to consist of decoupled hexagonal six-site clusters whose centers form a triangular superlattice (Fig.~\ref{fig:cluster}). This choice of cluster is sufficient to study N\'eel AF order, which is probed by adding to $H'$ the symmetry-breaking term
\begin{align}\label{weiss}
H_M'=M_z'\left(\sum_{i\in A}(n_{i\uparrow}-n_{i\downarrow})-\sum_{i\in B}(n_{i\uparrow}-n_{i\downarrow})\right),
\end{align}
where $A$ and $B$ correspond to sites within the cluster that belong to the two sublattices of the honeycomb lattice, and $M_z'$ is the Weiss field. In addition to $M_z'$, we also treat the chemical potential $\mu'$ of the cluster as a variational parameter. For a given value of the physical chemical potential $\mu$, which is chosen to maintain the system at half filling, $M_z'$ and $\mu'$ are used as variational parameters to extremize the Potthoff functional (\ref{Potthoff4}). It is necessary to consider the cluster chemical potential $\mu'$ as a variational parameter to ensure thermodynamic consistency, i.e., that the electronic density $n$ calculated from the trace of the Green's function $\Gv$ matches that obtained from the thermodynamic relation $n=-\partial\Omega/\partial\mu$, where $\Omega$ is the grand potential obtained from the Potthoff functional at its stationary point.\cite{aichhorn2006}

\section{VCA: Numerical results}
\label{sec:results}

\begin{figure}[t]
\includegraphics[scale = 0.9]{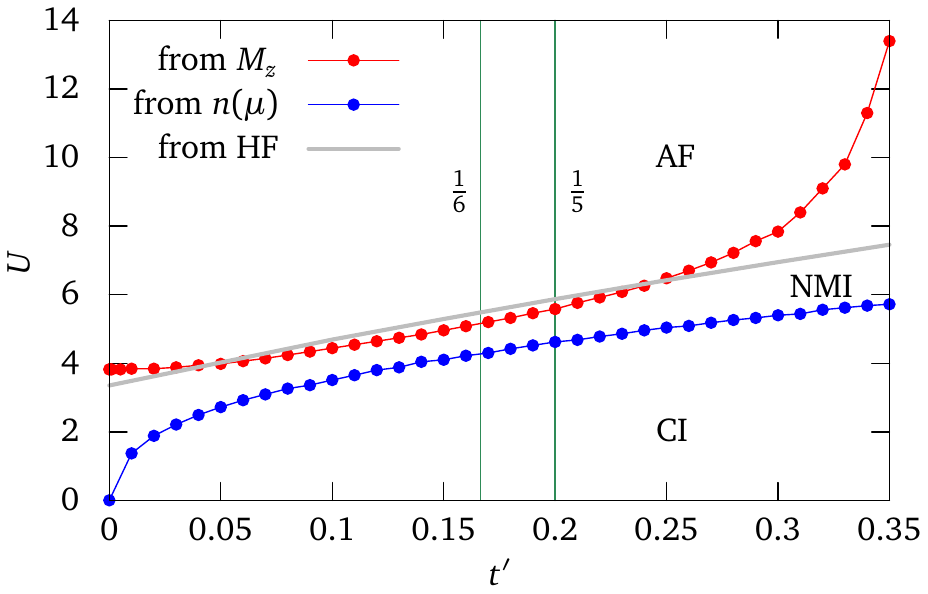}
\caption{(Color online) Ground-state phase diagram of the half-filled spinful Haldane-Hubbard model in the $U$-$t'$ plane for $\phi=\pi/2$, obtained in VCA. CI: Chern insulator, NMI: nonmagnetic insulator, AF: N\'{e}el antiferromagnetic insulator. The CI, NMI, and AF phases all have a nonzero one-particle (charge) gap. The CI-NMI phase boundary (blue circles) corresponds to a closing of the one-particle gap as determined from the one-particle density of states or the dependence of the electron density $n$ on the chemical potential $\mu$ (both methods closely agree). Shown for comparison, the solid gray line is the direct CI-AF transition found in the Hartree-Fock (HF) study of Ref.~\onlinecite{zheng2015}. The thin vertical green lines at $t'=1/6,1/5$ are cuts through the phase diagram across which various quantities are plotted in following figures.}
\label{fig:Utp}
\end{figure}

\begin{figure}[t]
\includegraphics[scale = 0.9]{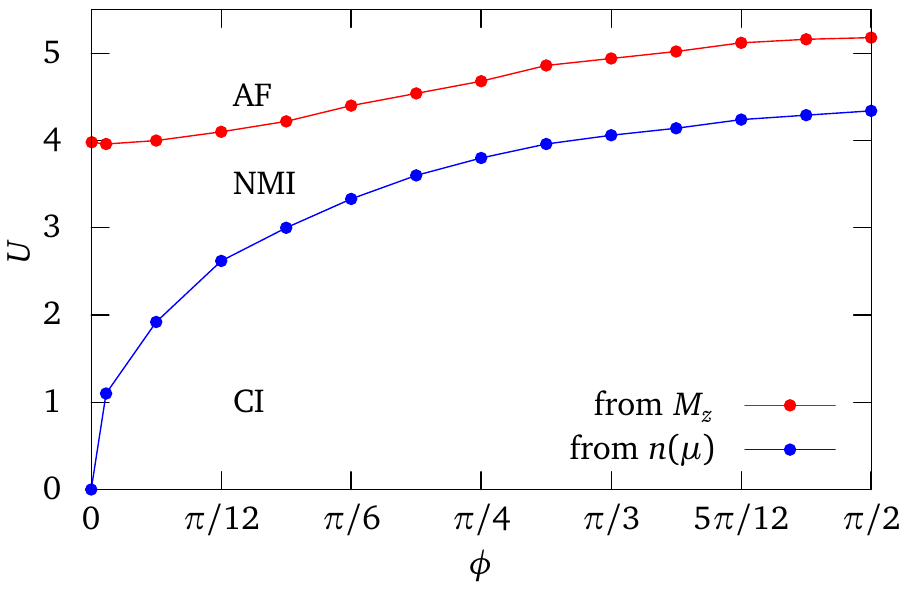}
\caption{(Color online) Ground-state phase diagram of the half-filled spinful Haldane-Hubbard model in the $U$-$\phi$ plane for $t'=1/6$, obtained in VCA. Phase boundaries are determined in the same way as in Fig.~\ref{fig:Utp}.}
  \label{fig:Uphi}
\end{figure}

\begin{figure}[t]
\includegraphics[width=\hsize]{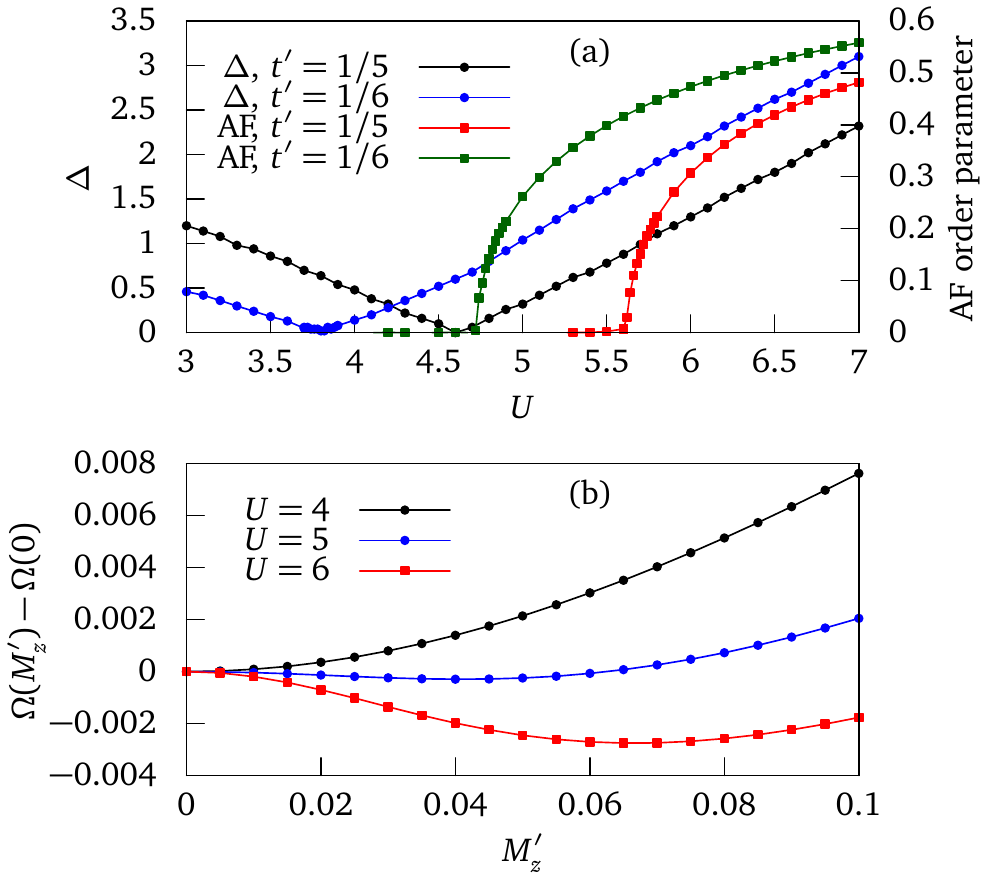}
\caption{(Color online) Quantum phase transitions in the half-filled spinful Haldane-Hubbard model, obtained in VCA. (a) One-particle gap $\Delta$ and AF order parameter as a function of $U$ for $t'=1/5$, $\phi=\pi/2$ and $t'=1/6$, $\phi=0.8$. (b) Potthoff functional $\Omega$ as a function of the Weiss field $M_z'$ for $t'=1/6$ and $\phi=0.8$.}
  \label{fig:qpt}
\end{figure}

We used VCA to determine the ground-state phase diagram of the half-filled spinful HH model in the $U$-$t'$ plane (Fig.~\ref{fig:Utp})  and in the $U$-$\phi$ plane (Fig.~\ref{fig:Uphi}). We find a correlated Chern insulator (CI) at small $U$, a N\'eel antiferromagnet (AF) at large $U$, and a nonmagnetic insulator (NMI) at intermediate $U$, sandwiched between the CI and AF phases. The AF order parameter (staggered magnetization), defined as the expectation value of the operator multiplying the Weiss field $M_z'$ in Eq.~(\ref{weiss}), is nonzero only in the AF phase and vanishes in the CI and NMI phases. All three phases have a nonzero one-particle gap and are thus insulating. At $t'=0$, the model reduces to the conventional Hubbard model with nearest-neighbor hopping on the honeycomb lattice. Although VCA finds a nonzero gap for all $U>0$, we know from large-scale QMC simulations\cite{sorella2012} that at $t'=0$ the gap remains zero for $U<3.869$, corresponding to a correlated semimetal. Above that critical value N\'eel AF order develops and a gap opens. There exists a finite critical $U$ for the opening of a gap at $t'=0$ because the $U=0$ low-energy spectrum of the model contains massless Dirac fermions that are protected by a combination of inversion and time-reversal symmetries. Unless they are broken spontaneously, those symmetries prevent the occurrence  of mass terms for the Dirac fermions.\cite{semenoff1984,haldane1988} At half filling, the chemical potential for the noninteracting problem is at the Dirac point, the density of states vanishes and there is a finite threshold value of $U$ for symmetry-breaking instabilities. Although VCA is unable to capture the gaplessness of the semimetallic region at $t'=0$, the onset of AF order is predicted correctly, with a critical $U_c=3.82$ at $t'=0$ very close to the QMC value $U_c=3.869$. Another cluster method, namely CDMFT, captures the gaplessness of the semimetallic region, and will be used in Sect.~\ref{sec:cdmft} to complement the VCA results presented in this section.

For $t'\neq 0$ and $\phi\neq 0,\pi$, the next-nearest-neighbor hopping term breaks time-reversal symmetry explicitly, and the Dirac fermions are gapped out already in the $U=0$ limit, corresponding to a noninteracting CI. Because it is gapped, the noninteracting CI evolves smoothly into a correlated CI upon increasing $U$. However, the one-particle gap decreases upon increasing $U$, and eventually closes and reopens at some critical value of $U$ that is $t'$ and $\phi$ dependent [Fig.~\ref{fig:qpt}(a)]. It appears to be linear in $U$ near the transition. The gap is calculated in two ways: from the one-particle density of states and from the dependence of $n$ on the chemical potential $\mu$, that is, from the compressibility. Both methods closely agree, and only one curve is shown. The closing of the one-particle gap occurs before the onset of magnetic order, unlike what is found in Hartree-Fock studies of the same Hamiltonian.\cite{zheng2015,arun2015} Furthermore, there are quantitative differences between ours and the Hartree-Fock result regarding the exact location of the AF phase boundary (Fig.~\ref{fig:Utp}). In particular, for $t'\gtrsim 0.25$ the AF phase boundary in VCA is pushed up to higher values of $U$ compared to the mean-field result. From the point of view of the effective spin model obtained from the HH model (\ref{H}) in the large-$U$ limit, it is natural to expect that the AF phase boundary would be pushed up in $U$ by a nonzero $t'$, which generates next-nearest-neighbor interaction terms that frustrate N\'eel AF order.\cite{hickey2015,hickey2015b} Given that it ignores the disordering effect of quantum fluctuations, conventional mean-field theory is known to overestimate the stability of magnetically ordered states and underestimate the effects of frustration. It is thus not surprising that the region of stability of AF order shrinks in VCA compared to the Hartree-Fock result.

\begin{figure}[t]
\includegraphics[scale = 0.9]{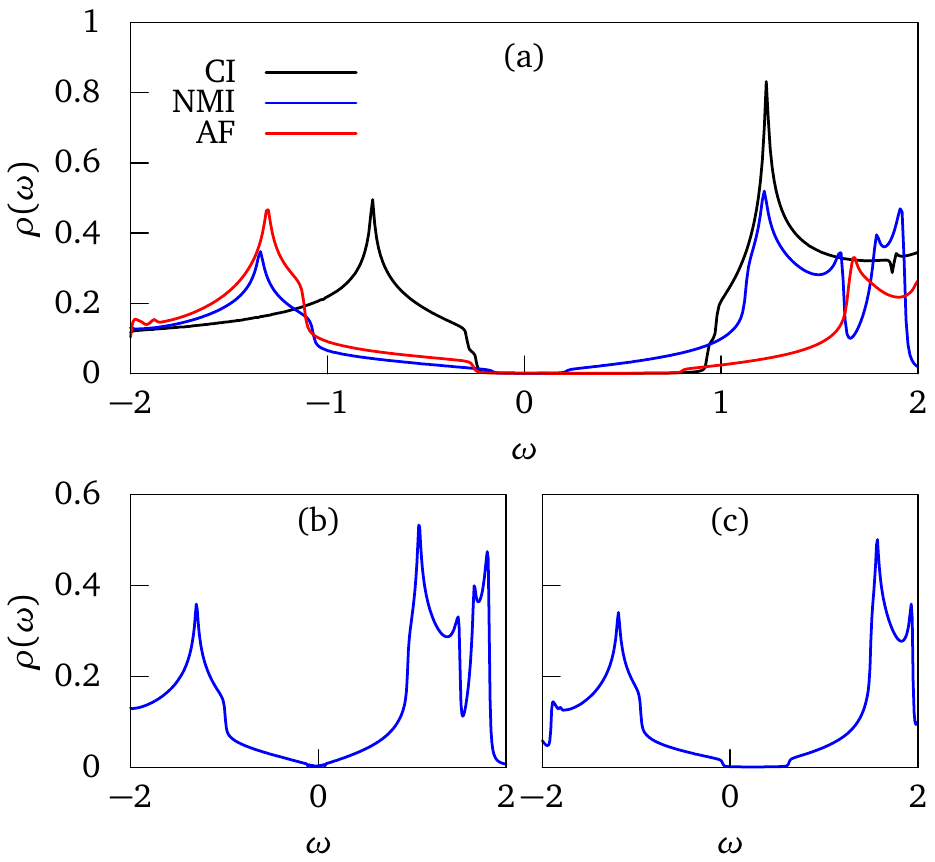}
\caption{(Color online) One-particle density of states $\rho(\omega)$ obtained in VCA for $t'=1/6$, $\phi=0.8$, as a function of frequency $\omega$ in (a) the correlated CI ($U=1$), NMI ($U=4.3$), and AF ($U=5$) phases; (b) at the CI-NMI transition ($U=3.81$); and (c) at the NMI-AF transition ($U=4.72$).}
  \label{fig:dos}
\end{figure}
%

In Fig.~\ref{fig:qpt}(a) we also plot the AF order parameter as a function of $U$, demonstrating that the NMI-AF transition is continuous. The AF order parameter is calculated from the CPT Green's function (\ref{GCPT}) at the stationary point of the Potthoff functional $\Omega$, i.e., with the values of $\mu'$ and $M_z'$ that extremize $\Omega$. In Fig.~\ref{fig:qpt}(b) we plot the Potthoff functional as a function of the AF Weiss field $M_z'$, at the value of the cluster chemical potential $\mu'$ that extremizes $\Omega$. This again clearly demonstrates the continuous nature of the NMI-AF transition. We find that at the stationary point, $\Omega$ is a minimum as a function of $M_z'$ but a maximum as a function of $\mu'$, as is often the case for models of correlated electrons.\cite{senechal2008}

VCA also allows one to calculate the one-particle density of states $\rho(\omega)$ (Fig.~\ref{fig:dos}) from the CPT Green's function (\ref{GCPT}) at the stationary point of the Potthoff functional. In Fig.~\ref{fig:dos}(a) we plot the density of states in the CI, NMI, and AF phases, which clearly displays a gap around $\omega=0$. Fig.~\ref{fig:dos}(b) and (c) show the density of states at the CI-NMI and NMI-AF transitions, respectively. At the CI-NMI transition, the density of states is roughly linear in frequency near the gap-closing point. Unlike the CI-NMI transition, the NMI-AF transition is not accompanied by a closing of the one-particle gap. In fact, as Fig.~\ref{fig:qpt}(a) shows, there is virtually no signature of the onset of magnetic order in the one-particle gap. This is very different from the mean-field picture, in which the onset of N\'eel AF order is immediately accompanied by a reduction of the one-particle gap.\cite{zheng2015}

\begin{figure}[t]
\includegraphics[scale = 0.9]{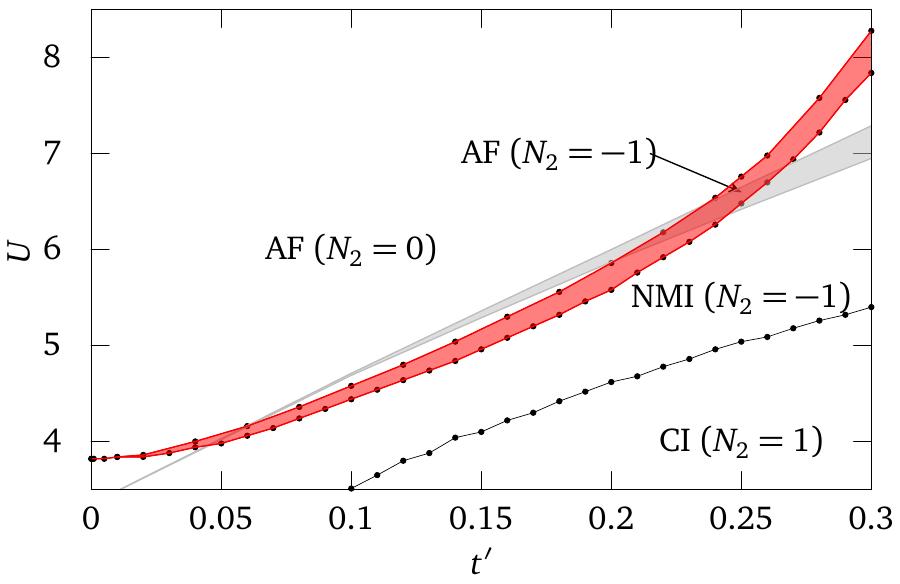}
\caption{(Color online) Generalized Chern number $N_2$ in the $U$-$t'$ plane for $\phi=\pi/2$, computed from the one-particle CPT Green's function obtained in VCA. The AF phase of Fig.~\ref{fig:Utp} contains a narrow topologically nontrivial region (red region) near the AF-NMI phase boundary. We show for comparison the topologically nontrivial AF region (gray region) found in the Hartree-Fock study of Ref.~\onlinecite{zheng2015}. }
  \label{fig:GCN}
\end{figure}

The exact nature of the NMI phase is difficult to pinpoint in our VCA calculations. By the nature of the method itself, we are limited to computing one-particle properties. As a further diagnostic of the phase, we have computed a topological invariant known as the generalized Chern number,\cite{niu1985,VolovikBook,wang2010}
\begin{align}\label{N2}
N_2=\frac{1}{6}\epsilon_{\mu\nu\lambda}\tr&\int_{-\infty}^\infty d\omega\int\frac{d^2k}{(2\pi)^2}\nonumber\\
&\times \Gv^{-1}\partial_{k_\mu}\Gv\Gv^{-1}\partial_{k_\nu}\Gv\Gv^{-1}
\partial_{k_\lambda}\Gv,
\end{align}
where $\mu,\nu,\lambda$ take values 0, 1, and 2 with $k_0\equiv\omega$, $\epsilon_{\mu\nu\lambda}$ is the fully antisymmetric tensor in three dimensions, $\Gv$ is the imaginary-time one-particle Green's function (here taken to be the CPT Green's function), $\Gv^{-1}$ is its matrix inverse, and the trace is taken over the matrix indices of $\Gv$, which include spin and band indices. From a mathematical standpoint, a $N\times N$ Green's function matrix $\Gv(\b{k},\omega)$ should be viewed as a mapping from frequency-momentum space $\mathbb{R}\times T^2$ to the space $\mathrm{GL}(N,\mathbb{C})$ of $N\times N$ complex-valued matrices, and Eq.~(\ref{N2}) is known as the Cartan-Maurer integral invariant, an integer-valued topological invariant that expresses the third homotopy class $\pi_3[\mathrm{GL}(N,\mathbb{C})]\cong\mathbb{Z}$ of this mapping. 

Rather than evaluating the frequency integral in Eq.~(\ref{N2}), we have used the simplified expression for the generalized Chern number derived in Ref.~\onlinecite{wang2012}, which only requires the knowledge of the Green's function at zero frequency:
\begin{equation}
N_2 = \int \frac{d^2k}{2\pi} \mathcal{F}_{xy}(\b{k}), \qquad
\mathcal{F}_{xy}(\b{k}) = \frac{\partial\mathcal{A}_y}{\partial k_x} - \frac{\partial \mathcal{A}_x}{\partial k_y},
\end{equation}
where
\begin{equation}
\mathcal{A}_j(\b{k}) = -i\kern-2ex\sum_{\alpha,\mu_\alpha(\b{k})>0} \langle\b{k},\alpha|\partial_{k_j}|\b{k},\alpha\rangle,
\end{equation}
and the $|\b{k},\alpha\rangle$ are the eigenvectors of the Green's function matrix $\Gv(\b{k},\omega=0)$ with positive eigenvalues $\mu_\alpha(\b{k})$ ($\Gv$ is a $4\times 4$ matrix because of the two bands and the two spin projections).
The practical computation of $N_2$ is done using the method proposed in Ref.~\onlinecite{Fukui:2005fk}, which
yields quantized (integer) values up to double precision accuracy ($10^{-16}$).

In Fig.~\ref{fig:GCN}, we illustrate the values taken by $N_2$ in the $U$-$t'$ plane. The gap-closing CI-NMI transition is accompanied by a topological transition at which $N_2$ changes from $+1$ in the CI phase to $-1$ in the NMI phase: thus the NMI phase is topologically nontrivial. The greater part of the AF phase has a vanishing generalized Chern number and is thus topologically trivial, except for a narrow sliver near the NMI-AF transition where $N_2=-1$ (red region in Fig.~\ref{fig:GCN}). To be precise, this topological AF region is bounded from below by the NMI-AF transition line of Fig.~\ref{fig:Utp} and from above by a topological transition at which $N_2$ changes by one. Similar results hold in the $U$-$\phi$ plane of Fig.~\ref{fig:Uphi}: there is a narrow topological AF region just above the NMI-AF phase boundary, but the one-particle gap there is too small to allow for an accurate determination of the location of the topological transition. For comparison, we also plot in Fig.~\ref{fig:GCN} the topological AF region found in the Hartree-Fock study of Ref.~\onlinecite{zheng2015} (gray region). Besides the different position of the transition lines, the latter study finds no NMI region and the topological AF region is sandwiched between the CI and the topologically trivial AF insulator; furthermore, the topological AF region has $N_2=1$ (per spin) as in the CI rather than $N_2=-1$.

\begin{figure}[t]
\includegraphics[width=\columnwidth]{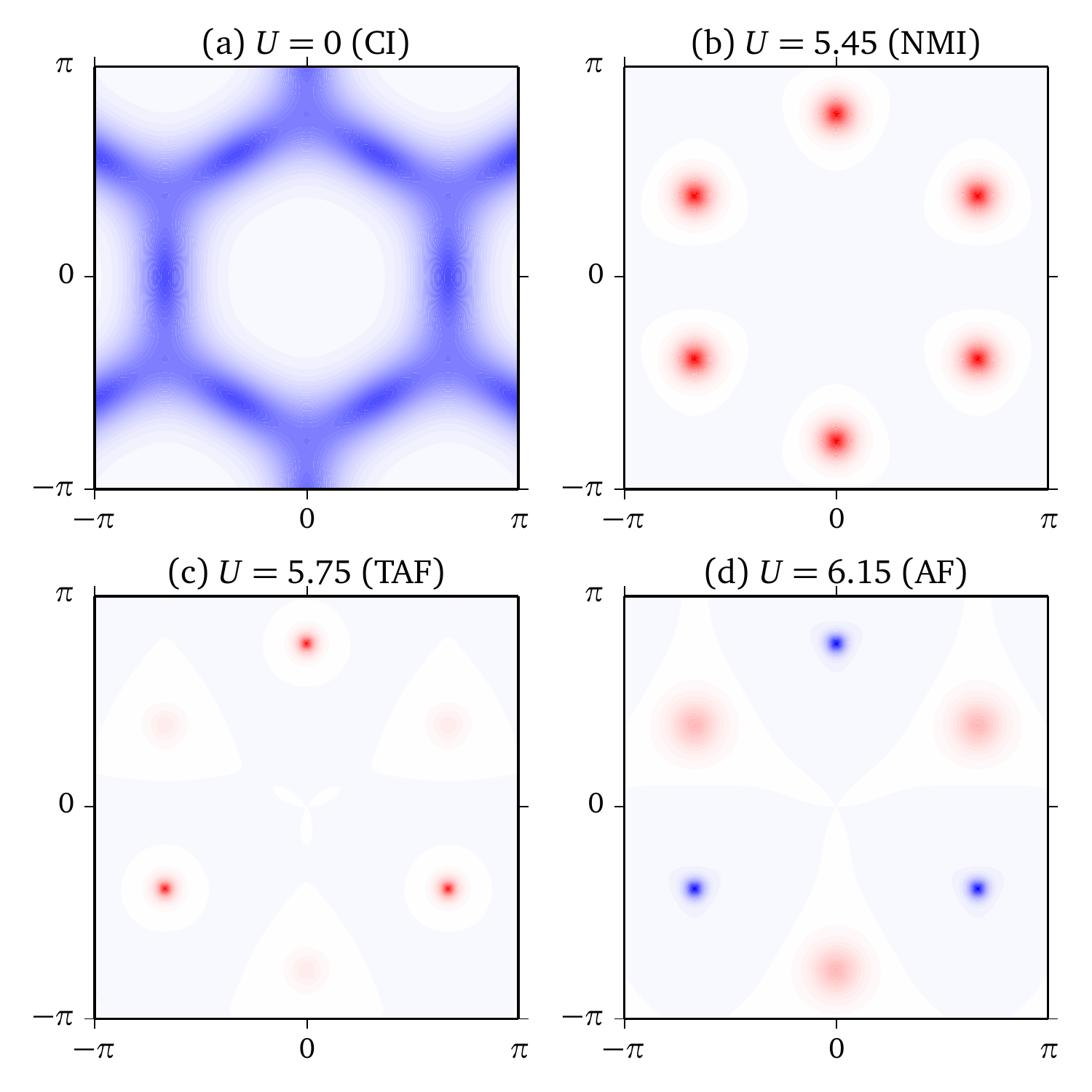}
\caption{(Color online) Berry curvature as a function of wavevector for $t'=0.2$, $\phi=\pi/2$ and four values of $U$ associated with the following phases obtained in VCA: (a) CI, (b) NMI, (c) topological AF, and (d) nontopological AF. Blue means positive, red means negative. Only the spin-up contribution is shown. The spin-down contribution is obtained by inverting with respect to the origin, but the two spin contributions to $N_2$ are equal.}
\label{fig:berry_vca}
\end{figure}

In Fig.~\ref{fig:berry_vca}, we show the Berry curvature $\mathcal{F}_{xy}(\b{k})$ as a function of wavevector at $t'=0.2$, $\phi=\pi/2$ and four values of $U$, corresponding to four different phases: CI, NMI, topological AF, and nontopological AF.
Whereas the change from $N_2=1$ to $N_2=-1$ at the CI-NMI transition is sudden, since we are then going through a gapless point, the passage from NMI to AF has a gradual effect on the Berry curvature map, weakening the contribution around the Dirac point $\b{K}$ and strengthening that from $\b{K}'$. At some value of $U$ in the AF phase, the contribution from $\b{K}'$ changes sign, the net contribution abruptly goes to zero and we fall into the topologically trivial AF phase.

How should one interpret these results for the generalized Chern number? First of all, the correlated CI at $U>0$ is adiabatically connected to the noninteracting CI at $U=0$. Since $N_2$ reduces to the single-particle Chern number in the noninteracting limit, and since the single-particle Chern number measures the Hall conductivity (here, per spin) in units of $e^2/h$,\cite{thouless1982} in the correlated CI phase also $N_2$ measures the Hall conductivity in units of $e^2/h$. However, this does not necessarily mean that the NMI and topological AF regions have quantized Hall conductivity $-2e^2/h$. The CI and NMI regions are separated by a phase transition, and the NMI and topological AF regions occur at a finite interaction strength $U>0$. In the presence of interactions, $N_2$ is not generally equal to the Hall conductivity.\cite{gurarie2013} The only statement one can safely make is that if one considers an interface between two semi-infinite CI and NMI regions, the difference between the number of gapless interface states and the number of zeros of the Green's function at the interface should equal the difference in generalized Chern numbers across the interface, namely four (accounting for the twofold spin degeneracy).\cite{gurarie2011} Therefore the NMI phase has a topological character, but it does not necessarily have a nonzero quantized Hall conductivity.

\section{CDMFT: Numerical results}
\label{sec:cdmft}

As a complement to our VCA results, we have also studied the same system using CDMFT.\cite{Lichtenstein:2000vn,Kotliar:2001} We will not provide a review of the method here, but rather refer the reader to the literature.\cite{Bolech:2003ye, Kancharla:2008vn, Senechal:2015rm} The method proceeds like CPT and VCA: an effective model is solved on a small cluster, and the self-energy associated with that cluster is applied to the whole lattice.
However, the effect of the cluster's environment is not embodied in various Weiss fields residing on the cluster, but rather by a set of uncorrelated, additional orbitals hybridized with the cluster (the ``bath''). 
These bath orbitals have their own (possibly spin-dependent) energy levels ($\varepsilon_{i\s}$) and are hybridized with the cluster sites with amplitudes $\theta_{i\s}$.
The bath parameters ($\varepsilon_{i\s},\theta_{i\s}$) are determined by a self-consistency condition.
With an ED solver, the computational complexity is determined by the total number of orbitals (cluster plus bath) and a compromise must be made between the number of bath orbitals and the number of sites in the cluster. A better resolution in the time domain, and therefore a better rendering of spectral properties, is generally obtained by increasing the number of bath orbitals at the expense of cluster sites. But this in turn deteriorates the spatial resolution of the method.

\begin{figure}[t]
\includegraphics[width=\hsize]{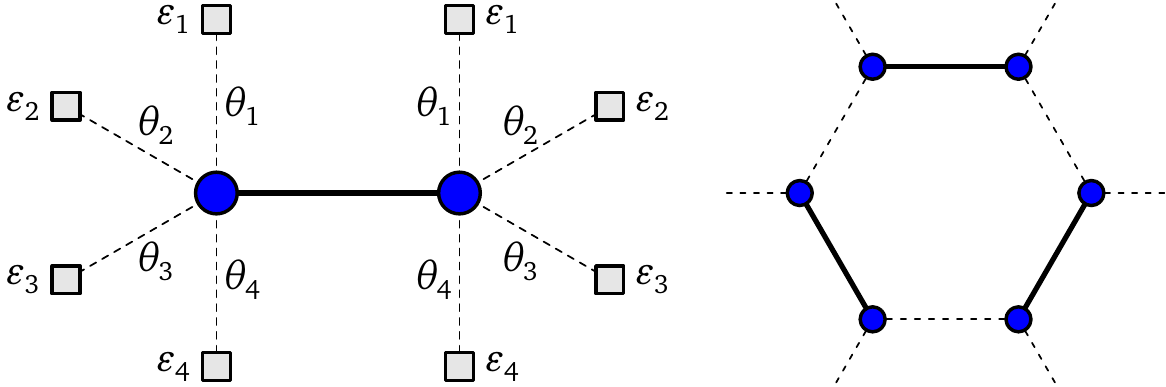}
\caption{(Color online) Left: Cluster-bath system used in CDMFT. The cluster contains two sites (blue symbols) forming the unit cell of the model. The bath sites are indicated by gray squares. They have energies $\varepsilon_i$ ($i=1,\ldots,4$) and are hybridized with the cluster sites as indicated (dashed lines) with hopping amplitudes $\theta_i$. Right: Arrangement of this cluster to form a repeated pattern.}
\label{fig:cdmft_cluster}
\end{figure}

\begin{figure}[t]
\includegraphics[scale=0.9]{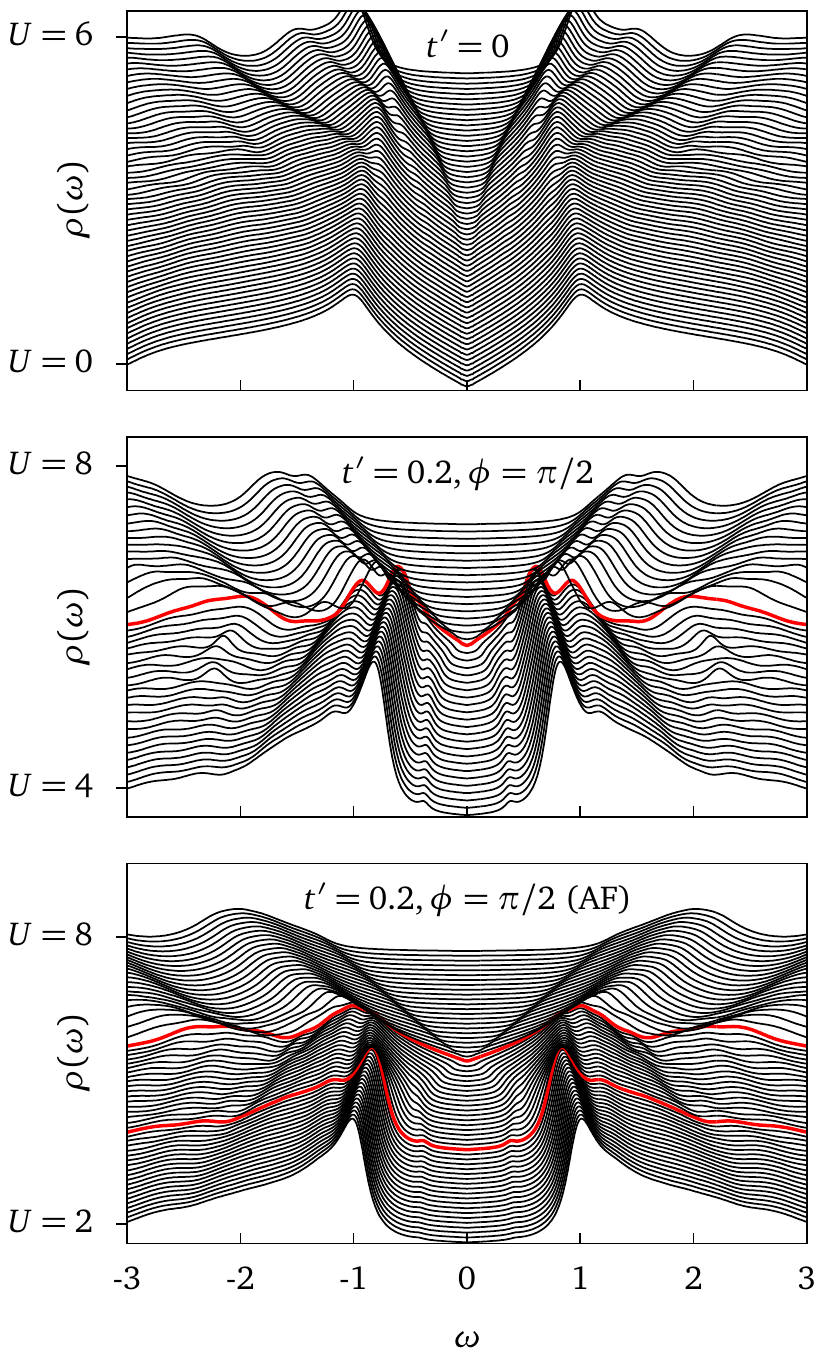}
\caption{(Color online) One-particle density of states $\rho(\omega)$ computed from CDMFT for a range of values of $U$, shifted for clarity. On the top panel ($t'=0$), we observe a transition at $U=5.6$ from the semimetal to the Mott insulator. On the middle panel ($t'=0.2$, $\phi=\pi/2$), antiferromagnetism is suppressed by hand and a transition occurs from the CI to the Mott insulator at $U=6.1$. On the bottom panel, antiferromagnetism is allowed to develop and two transitions occur: a CI to a topological AF at $U=3.9$, followed by a topological transition to an ordinary AF at $U=5.8$.}
\label{fig:dos_cdmft}
\end{figure}

\begin{figure}[t]
\includegraphics[scale=0.9]{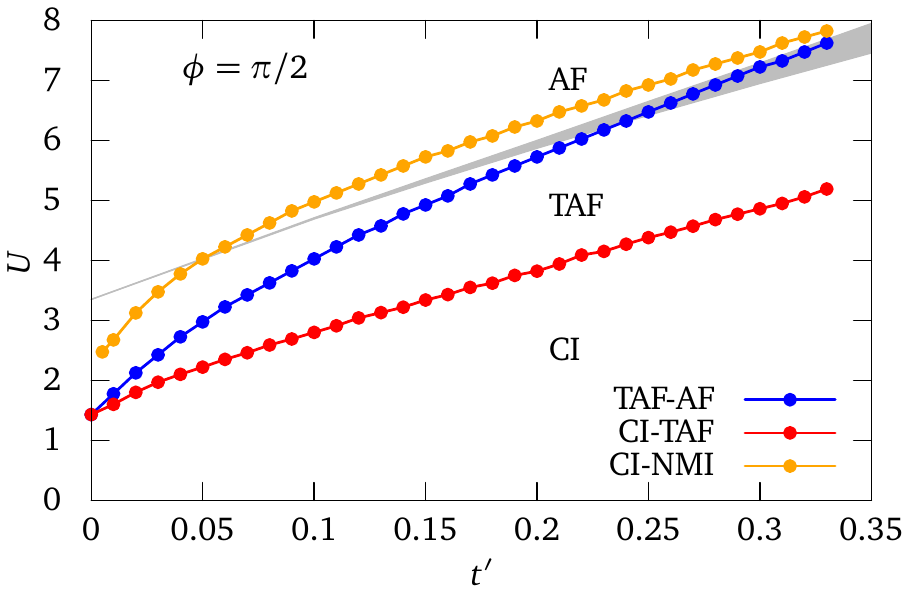}
\caption{(Color online) Ground-state phase diagram of the half-filled spinful Haldane-Hubbard model in the $U$-$t'$ plane for $\phi=\pi/2$, according to CDMFT computations based on the cluster-bath system of Fig.~\ref{fig:cdmft_cluster}. CI: Chern insulator, TAF: Topological antiferromagnetic insulator, AF: ordinary antiferromagnetic insulator. The CI and TAF phases have topological charge $N_2=1$. The CI-TAF and TAF-AF phase boundaries are shown in red and blue, respectively. Also shown is the boundary between the CI and the NMI Mott phase when suppressing antiferromagnetism.
In that case the boundary also marks a topological change: the Mott phase has $N_2=-1$. Again, the topologically nontrivial AF region found in the Hartree-Fock study of Ref.~\onlinecite{zheng2015} is shown (gray area).}
\label{fig:cdmft_phase_diagram}
\end{figure}

\begin{figure}[t]
\includegraphics[width=\columnwidth]{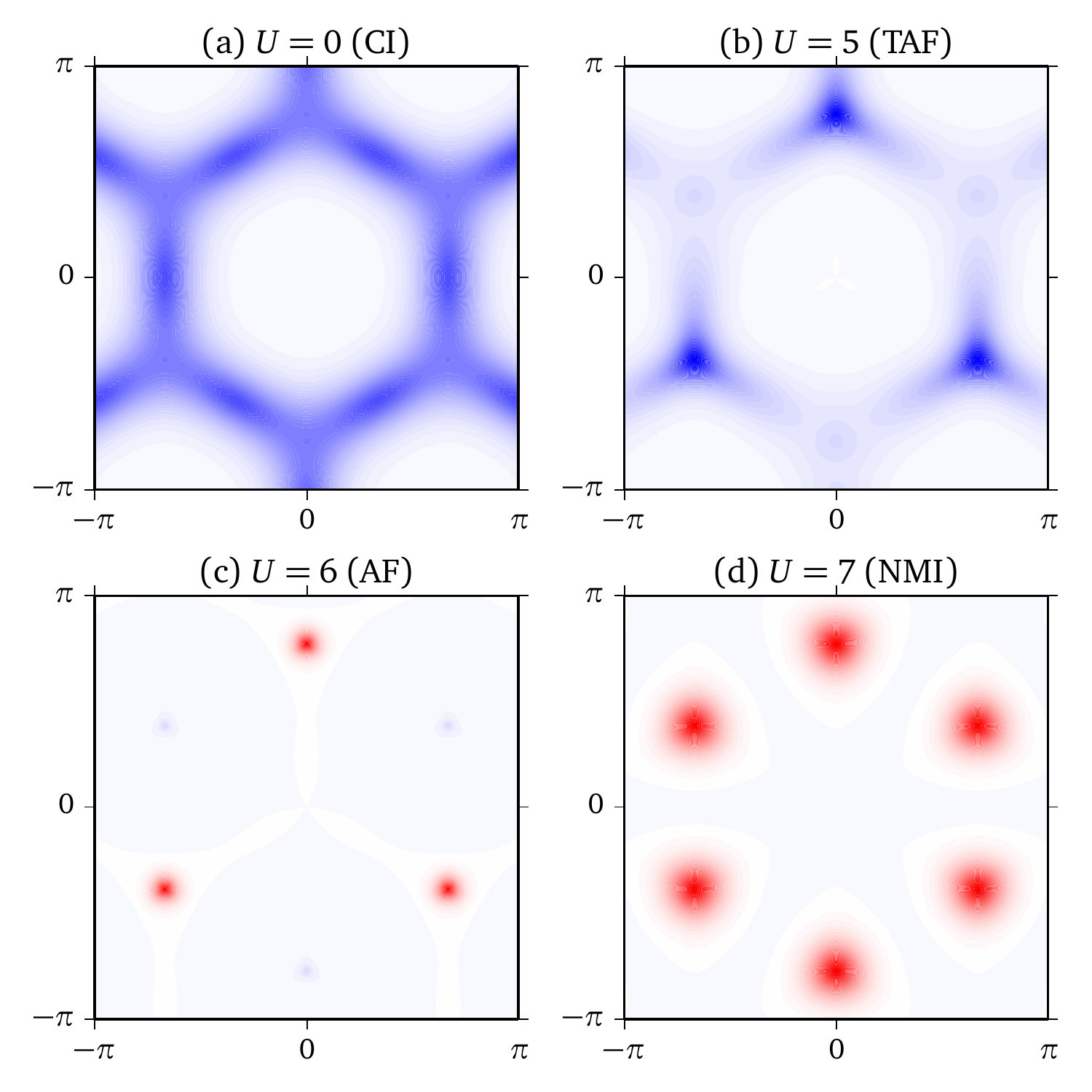}
\caption{(Color online) Berry curvature as a function of wavevector for $t'=0.2$, $\phi=\pi/2$ and four values of $U$ associated with the following phases obtained in CDMFT: (a) CI, (b) TAF, (c) AF, and (d) NMI (normal phase). Blue means positive, red means negative. Only the spin-up contribution is shown. The spin-down contribution is obtained by inverting with respect to the origin, but the two spin contributions to $N_2$ are equal.}
\label{fig:berry_cdmft}
\end{figure}

In order to study the spectral properties of model \eqref{H}, we will use the cluster-bath system illustrated in Fig.~\ref{fig:cdmft_cluster}, which contains only two cluster sites (the unit cell), and eight bath sites.
The two-site clusters are arranged to form a hexagon, as indicated on the right, and these hexagons are repeated just like the VCA cluster of Fig.~\ref{fig:cluster}.
The computation can be carried out by suppressing antiferromagnetism, i.e., by assuming that the bath parameters are independent of spin and identical for the two sublattices, or by allowing antiferromagnetism to develop with spin and sublattice-dependent bath parameters. 
The density of states $\rho(\omega)$ for two values of $t'$ ($0$ and $0.2$) is shown in Fig.~\ref{fig:dos_cdmft}.
The top panel ($t'=0$) shows the transition from the semimetallic state at weak coupling to the Mott insulator at strong coupling, when antiferromagnetism is suppressed. The V-like line shape of the semimetallic state is correctly reproduced by CDMFT, and the Mott transition is followed by a gap that increases linearly with $U$ thereafter. With the chosen cluster-bath system, no sharp transition occurs as a function of $U$ at $t'=0$: it is a crossover, albeit a rather well-defined one. However, a sharp transition is observed for $t'>0.02$.
In CDMFT, an additional signature of the Mott transition is the abrupt change in behavior of the bath parameters $(\varepsilon_i,\theta_i)$, sometimes with hysteretic behavior.
The middle panel of Fig.~\ref{fig:dos_cdmft} shows the density of states at $t'=0.2$ and $\phi=\pi/2$.
The difference lies in the weak coupling phase, which is a gapped CI. The gap vanishes at the Mott transition (red curve), and the invariant $N_2$ goes suddenly from $+1$ to $-1$ across the transition, indicating that we are entering the NMI phase found in VCA.
Finally, in the bottom panel, antiferromagnetism was allowed to develop. In that case, two transitions occur, indicated by red curves: the first one, at $U\approx 3.8$, from a CI to a topological AF, without closure of the gap (the invariant $N_2$ remains equal to $+1$). A second transition occurs at $U\approx5.7$, through a gapless point, when $N_2$ suddenly drops to zero, towards an ordinary, nontopological AF. In all these spectral plots, a Lorentzian broadening is added to each peak of the spectral function $A(\b{k},\omega)$, and this broadening has been set to increase with $|\omega|$, because high-frequency features obtained from an ED solver are not as accurate as low-frequency ones. This removes the sharpness of gap edges.  

Figure \ref{fig:cdmft_phase_diagram} shows where these transitions occur in the $U$-$t'$ plane, for a fixed value $\phi=\pi/2$ of the phase. This is the CDMFT version of Fig.~\ref{fig:Utp}.
The orange curve shows the Mott CI-NMI transition found when antiferromagnetism is suppressed.
The red curve is the first transition, from CI to topological AF (TAF), and the blue curve is the second transition, towards a nontopological AF. The Hartree-Fock result of Ref.~\onlinecite{zheng2015} is again shown, in gray.
The topological AF phase disappears at $t'=0$. Beyond $t'=0.33$, the $\b{Q}=0$ N\'eel phase is likely no longer the correct magnetic order to probe, as hinted at by Hartree-Fock computations.
Once in the AF phase, i.e., when antiferromagnetism is not suppressed, no hint of the Mott transition occurs across the (orange) Mott line: the NMI phase is completely preempted by the antiferromagnetic state.

The critical value $U_c$ for the onset of antiferromagnetism at $t'=0$ is $U_c=1.45$. This is obviously incorrect, and must be attributed to the small size of the cluster used in CDMFT (two sites). In cluster methods, small clusters tend to exaggerate the effect of $U$. It is thus conceivable that larger clusters might push the onset of AF order to higher values of $U$ and expose a stable region of NMI phase as in the VCA results.

In order to confirm the mapping between the phases observed in VCA and CDMFT, we show in Fig.~\ref{fig:berry_cdmft} the Berry curvature map in the first Brillouin zone for the solutions found in CDMFT. This is to be compared with the same plots in Fig.~\ref{fig:berry_vca}, obtained in VCA. In both cases the CI phase is characterized by a positive Berry curvature concentrated along the Brillouin zone edges and the NMI phase by a negative curvature concentrated around the Dirac points. The difference lies in the respective positions of these phases with respect to the magnetic phases. The latter display the expected symmetry breaking between the two inequivalent Dirac points (only the spin-up density of states is shown).

\section{Conclusion}
\label{sec:concl}

We have investigated the ground-state phase diagram of the half-filled spinful HH model with a combination of quantum cluster methods: CPT, VCA, and CDMFT. In agreement with previous mean-field studies,\cite{zheng2015} in both VCA and CDMFT we find a topologically trivial N\'eel AF at large $U$ and a correlated CI at small $U$. Here we define phases as being topologically nontrivial if their generalized Chern number $N_2$, defined as a winding number of the one-particle Green's function (here approximated as the CPT Green's function), is nonzero. For intermediate interactions, in both VCA and CDMFT we find topologically nontrivial N\'eel AF phases. This is also in agreement with previous studies, although the precise ordering of these phases and their value of $N_2$ depends on the method. To the difference of previous studies however, we also find a topologically nontrivial NMI phase in the intermediate interaction regime. Whether this phase is stabilized as the actual ground state in our calculations or is preempted by the onset of AF order depends on the method used.

Given that quantum cluster methods are essentially designed to determine one-particle properties of correlated systems, our study does not allow us to precisely pinpoint the nature of the NMI phase. A tantalizing possibility is that this phase could be a fractionalized topological phase such as the CSL\cite{he2011,hickey2015} or the CI*,\cite{maciejko2013,prychynenko2014} which have been predicted in this model by slave-particle mean-field approaches. In particular, to the difference of the CSL, the CI* has a nonzero quantized Hall conductivity $\sigma_{xy}=\pm 2e^2/h$,\cite{maciejko2013} which would be consistent with the value $N_2=-1$ found in the NMI phase---that is, if $N_2$ does happen to coincide with the Hall conductivity (per spin) in this case. To determine unambiguously whether the NMI phase corresponds to the CSL or the CI*, one would first have to show the existence of intrinsic topological order in the NMI phase via the demonstration of topological ground-state degeneracy $D$ on the torus and/or the fractional statistics of excitations. The CSL would correspond to $D=2$ and semionic excitations, while the CI* would have $D=4$ and excitations with semionic, antisemionic, and bosonic statistics. However, the quantum cluster methods used here do not allow us to determine these properties, which require the knowledge of the full many-body ground-state wave function,\cite{zhang2012} and one must use methods such as ED or the density-matrix renormalization group (DMRG).

From the point of view of effective spin models valid in the strong Mott regime $U\gg 1$,\cite{hickey2015,hickey2015b} the NMI phase found here would correspond to the weak Mott regime $U\sim 1$ in which sizable ring-exchange spin interactions induced by a small charge gap could frustrate magnetic order and stabilize exotic quantum disordered phases.\cite{motrunich2005,lee2005} Of course, a more mundane possibility is that the NMI is simply adiabatically connected to a (topological) band insulator without fractionalization or topological order. The NMI could then be considered a topological Mott insulator in the sense of Ref.~\onlinecite{raghu2008}. In general one should also consider the possibility of other types of magnetic order besides N\'eel order, but in previous studies these occur either in the $U\gg 1$ limit\cite{hickey2015,hickey2015b} or for $t'\gtrsim 0.35$.\cite{zheng2015,arun2015}

Further numerical studies are clearly needed to resolve the difference between VCA and CDMFT predictions and fully elucidate the ground-state properties of the spinful HH model, especially in the regime of intermediate repulsion $U\sim 1$. In our opinion, by Occam's razor the most likely scenario is the CDMFT one, in which the NMI phase is preempted by the onset of conventional AF order. However, it is known that in CDMFT the critical $U$ for AF order will increase with cluster size, as small clusters have comparatively fewer links than sites and thus overestimate the effect of on-site interactions relative to inter-site hopping. The critical $U$ for the Mott transition, on the other hand, does not vary much with cluster size, that transition being more of a local phenomenon compared to the AF transition. Although not extremely likely, it is thus possible that the AF transition might be pushed beyond the Mott transition even in CDMFT, for larger clusters. Finally, it is possible that adding frustrating interactions to the HH model such as third neighbor hopping\cite{hickey2015b} might stabilize the NMI phase and realize the VCA scenario. As mentioned earlier, studies of models of interacting fermions where explicitly broken time-reversal and/or particle-hole symmetries preclude the use of powerful QMC methods are notoriously hard. Besides VCA and CDMFT however, other powerful numerical methods have been successfully applied recently to the study of models of correlated Chern insulators, such as the cellular dynamical impurity approximation\cite{faye2014} (CDIA) and DMRG.\cite{alba2015,grushin2015} It would be worthwhile to apply these methods to the study of the spinful HH model at half filling.

\acknowledgements

We acknowledge helpful discussions with G. Chen and R. Thomale. J.M. wishes to acknowledge the hospitality of the Kavli Institute for Theoretical Physics (KITP) where part of this research was carried out. J.W. was supported by the Chinese Scholarship Council (CSC). D.S. was supported by NSERC grant \#RGPIN-2015-05598. J.M. was supported by NSERC grant \#RGPIN-2014-4608, the Canada Research Chair Program (CRC), the Canadian Institute for Advanced Research (CIFAR), and the University of Alberta. Computational resources were provided by Compute Canada, Calcul Qu\'ebec, and WestGrid. This research was also supported in part by the National Science Foundation under Grant No.~NSF PHY11-25915 (KITP).

\bibliography{spinfulhh}

\begin{thebibliography}{87}%
\makeatletter
\providecommand \@ifxundefined [1]{%
 \@ifx{#1\undefined}
}%
\providecommand \@ifnum [1]{%
 \ifnum #1\expandafter \@firstoftwo
 \else \expandafter \@secondoftwo
 \fi
}%
\providecommand \@ifx [1]{%
 \ifx #1\expandafter \@firstoftwo
 \else \expandafter \@secondoftwo
 \fi
}%
\providecommand \natexlab [1]{#1}%
\providecommand \enquote  [1]{``#1''}%
\providecommand \bibnamefont  [1]{#1}%
\providecommand \bibfnamefont [1]{#1}%
\providecommand \citenamefont [1]{#1}%
\providecommand \href@noop [0]{\@secondoftwo}%
\providecommand \href [0]{\begingroup \@sanitize@url \@href}%
\providecommand \@href[1]{\@@startlink{#1}\@@href}%
\providecommand \@@href[1]{\endgroup#1\@@endlink}%
\providecommand \@sanitize@url [0]{\catcode `\\12\catcode `\$12\catcode
  `\&12\catcode `\#12\catcode `\^12\catcode `\_12\catcode `\%12\relax}%
\providecommand \@@startlink[1]{}%
\providecommand \@@endlink[0]{}%
\providecommand \url  [0]{\begingroup\@sanitize@url \@url }%
\providecommand \@url [1]{\endgroup\@href {#1}{\urlprefix }}%
\providecommand \urlprefix  [0]{URL }%
\providecommand \Eprint [0]{\href }%
\providecommand \doibase [0]{http://dx.doi.org/}%
\providecommand \selectlanguage [0]{\@gobble}%
\providecommand \bibinfo  [0]{\@secondoftwo}%
\providecommand \bibfield  [0]{\@secondoftwo}%
\providecommand \translation [1]{[#1]}%
\providecommand \BibitemOpen [0]{}%
\providecommand \bibitemStop [0]{}%
\providecommand \bibitemNoStop [0]{.\EOS\space}%
\providecommand \EOS [0]{\spacefactor3000\relax}%
\providecommand \BibitemShut  [1]{\csname bibitem#1\endcsname}%
\let\auto@bib@innerbib\@empty
\bibitem [{\citenamefont {Jotzu}\ \emph {et~al.}(2014)\citenamefont {Jotzu},
  \citenamefont {Messer}, \citenamefont {Desbuquois}, \citenamefont {Lebrat},
  \citenamefont {Uehlinger}, \citenamefont {Greif},\ and\ \citenamefont
  {Esslinger}}]{jotzu2014}%
  \BibitemOpen
  \bibfield  {author} {\bibinfo {author} {\bibfnamefont {G.}~\bibnamefont
  {Jotzu}}, \bibinfo {author} {\bibfnamefont {M.}~\bibnamefont {Messer}},
  \bibinfo {author} {\bibfnamefont {R.}~\bibnamefont {Desbuquois}}, \bibinfo
  {author} {\bibfnamefont {M.}~\bibnamefont {Lebrat}}, \bibinfo {author}
  {\bibfnamefont {T.}~\bibnamefont {Uehlinger}}, \bibinfo {author}
  {\bibfnamefont {D.}~\bibnamefont {Greif}}, \ and\ \bibinfo {author}
  {\bibfnamefont {T.}~\bibnamefont {Esslinger}},\ }\href
  {http://www.nature.com/nature/journal/v515/n7526/full/nature13915.html}
  {\bibfield  {journal} {\bibinfo  {journal} {Nature}\ }\textbf {\bibinfo
  {volume} {515}},\ \bibinfo {pages} {237} (\bibinfo {year}
  {2014})}\BibitemShut {NoStop}%
\bibitem [{\citenamefont {Haldane}(1988)}]{haldane1988}%
  \BibitemOpen
  \bibfield  {author} {\bibinfo {author} {\bibfnamefont {F.~D.~M.}\
  \bibnamefont {Haldane}},\ }\href
  {http://link.aps.org/doi/10.1103/PhysRevLett.61.2015} {\bibfield  {journal}
  {\bibinfo  {journal} {Phys. Rev. Lett.}\ }\textbf {\bibinfo {volume} {61}},\
  \bibinfo {pages} {2015} (\bibinfo {year} {1988})}\BibitemShut {NoStop}%
\bibitem [{\citenamefont {Chang}\ \emph {et~al.}(2013)\citenamefont {Chang},
  \citenamefont {Zhang}, \citenamefont {Feng}, \citenamefont {Shen},
  \citenamefont {Zhang}, \citenamefont {Guo}, \citenamefont {Li}, \citenamefont
  {Ou}, \citenamefont {Wei}, \citenamefont {Wang}, \citenamefont {Ji},
  \citenamefont {Feng}, \citenamefont {Ji}, \citenamefont {Chen}, \citenamefont
  {Jia}, \citenamefont {Dai}, \citenamefont {Fang}, \citenamefont {Zhang},
  \citenamefont {He}, \citenamefont {Wang}, \citenamefont {Lu}, \citenamefont
  {Ma},\ and\ \citenamefont {Xue}}]{chang2013}%
  \BibitemOpen
  \bibfield  {author} {\bibinfo {author} {\bibfnamefont {C.-Z.}\ \bibnamefont
  {Chang}}, \bibinfo {author} {\bibfnamefont {J.}~\bibnamefont {Zhang}},
  \bibinfo {author} {\bibfnamefont {X.}~\bibnamefont {Feng}}, \bibinfo {author}
  {\bibfnamefont {J.}~\bibnamefont {Shen}}, \bibinfo {author} {\bibfnamefont
  {Z.}~\bibnamefont {Zhang}}, \bibinfo {author} {\bibfnamefont
  {M.}~\bibnamefont {Guo}}, \bibinfo {author} {\bibfnamefont {K.}~\bibnamefont
  {Li}}, \bibinfo {author} {\bibfnamefont {Y.}~\bibnamefont {Ou}}, \bibinfo
  {author} {\bibfnamefont {P.}~\bibnamefont {Wei}}, \bibinfo {author}
  {\bibfnamefont {L.-L.}\ \bibnamefont {Wang}}, \bibinfo {author}
  {\bibfnamefont {Z.-Q.}\ \bibnamefont {Ji}}, \bibinfo {author} {\bibfnamefont
  {Y.}~\bibnamefont {Feng}}, \bibinfo {author} {\bibfnamefont {S.}~\bibnamefont
  {Ji}}, \bibinfo {author} {\bibfnamefont {X.}~\bibnamefont {Chen}}, \bibinfo
  {author} {\bibfnamefont {J.}~\bibnamefont {Jia}}, \bibinfo {author}
  {\bibfnamefont {X.}~\bibnamefont {Dai}}, \bibinfo {author} {\bibfnamefont
  {Z.}~\bibnamefont {Fang}}, \bibinfo {author} {\bibfnamefont {S.-C.}\
  \bibnamefont {Zhang}}, \bibinfo {author} {\bibfnamefont {K.}~\bibnamefont
  {He}}, \bibinfo {author} {\bibfnamefont {Y.}~\bibnamefont {Wang}}, \bibinfo
  {author} {\bibfnamefont {L.}~\bibnamefont {Lu}}, \bibinfo {author}
  {\bibfnamefont {X.-C.}\ \bibnamefont {Ma}}, \ and\ \bibinfo {author}
  {\bibfnamefont {Q.-K.}\ \bibnamefont {Xue}},\ }\href
  {http://www.sciencemag.org/content/340/6129/167} {\bibfield  {journal}
  {\bibinfo  {journal} {Science}\ }\textbf {\bibinfo {volume} {340}},\ \bibinfo
  {pages} {167} (\bibinfo {year} {2013})}\BibitemShut {NoStop}%
\bibitem [{\citenamefont {Bestwick}\ \emph {et~al.}(2015)\citenamefont
  {Bestwick}, \citenamefont {Fox}, \citenamefont {Kou}, \citenamefont {Pan},
  \citenamefont {Wang},\ and\ \citenamefont {Goldhaber-Gordon}}]{bestwick2015}%
  \BibitemOpen
  \bibfield  {author} {\bibinfo {author} {\bibfnamefont {A.}~\bibnamefont
  {Bestwick}}, \bibinfo {author} {\bibfnamefont {E.}~\bibnamefont {Fox}},
  \bibinfo {author} {\bibfnamefont {X.}~\bibnamefont {Kou}}, \bibinfo {author}
  {\bibfnamefont {L.}~\bibnamefont {Pan}}, \bibinfo {author} {\bibfnamefont
  {K.~L.}\ \bibnamefont {Wang}}, \ and\ \bibinfo {author} {\bibfnamefont
  {D.}~\bibnamefont {Goldhaber-Gordon}},\ }\href
  {http://link.aps.org/doi/10.1103/PhysRevLett.114.187201} {\bibfield
  {journal} {\bibinfo  {journal} {Phys. Rev. Lett.}\ }\textbf {\bibinfo
  {volume} {114}},\ \bibinfo {pages} {187201} (\bibinfo {year}
  {2015})}\BibitemShut {NoStop}%
\bibitem [{\citenamefont {Chang}\ \emph {et~al.}(2015)\citenamefont {Chang},
  \citenamefont {Zhao}, \citenamefont {Kim}, \citenamefont {Zhang},
  \citenamefont {Assaf}, \citenamefont {Heiman}, \citenamefont {Zhang},
  \citenamefont {Liu}, \citenamefont {Chan},\ and\ \citenamefont
  {Moodera}}]{chang2015}%
  \BibitemOpen
  \bibfield  {author} {\bibinfo {author} {\bibfnamefont {C.-Z.}\ \bibnamefont
  {Chang}}, \bibinfo {author} {\bibfnamefont {W.}~\bibnamefont {Zhao}},
  \bibinfo {author} {\bibfnamefont {D.~Y.}\ \bibnamefont {Kim}}, \bibinfo
  {author} {\bibfnamefont {H.}~\bibnamefont {Zhang}}, \bibinfo {author}
  {\bibfnamefont {B.~A.}\ \bibnamefont {Assaf}}, \bibinfo {author}
  {\bibfnamefont {D.}~\bibnamefont {Heiman}}, \bibinfo {author} {\bibfnamefont
  {S.-C.}\ \bibnamefont {Zhang}}, \bibinfo {author} {\bibfnamefont
  {C.}~\bibnamefont {Liu}}, \bibinfo {author} {\bibfnamefont {M.~H.~W.}\
  \bibnamefont {Chan}}, \ and\ \bibinfo {author} {\bibfnamefont {J.~S.}\
  \bibnamefont {Moodera}},\ }\href
  {http://www.nature.com/nmat/journal/v14/n5/full/nmat4204.html} {\bibfield
  {journal} {\bibinfo  {journal} {Nature Mater.}\ }\textbf {\bibinfo {volume}
  {14}},\ \bibinfo {pages} {473} (\bibinfo {year} {2015})}\BibitemShut
  {NoStop}%
\bibitem [{\citenamefont {Greiner}\ \emph {et~al.}(2002)\citenamefont
  {Greiner}, \citenamefont {Mandel}, \citenamefont {Esslinger}, \citenamefont
  {{H\"{a}nsch}},\ and\ \citenamefont {Bloch}}]{greiner2002}%
  \BibitemOpen
  \bibfield  {author} {\bibinfo {author} {\bibfnamefont {M.}~\bibnamefont
  {Greiner}}, \bibinfo {author} {\bibfnamefont {O.}~\bibnamefont {Mandel}},
  \bibinfo {author} {\bibfnamefont {T.}~\bibnamefont {Esslinger}}, \bibinfo
  {author} {\bibfnamefont {T.~W.}\ \bibnamefont {{H\"{a}nsch}}}, \ and\
  \bibinfo {author} {\bibfnamefont {I.}~\bibnamefont {Bloch}},\ }\href
  {http://www.nature.com/nature/journal/v415/n6867/full/415039a.html}
  {\bibfield  {journal} {\bibinfo  {journal} {Nature}\ }\textbf {\bibinfo
  {volume} {415}},\ \bibinfo {pages} {39} (\bibinfo {year} {2002})}\BibitemShut
  {NoStop}%
\bibitem [{\citenamefont {{K\"{o}hl}}\ \emph {et~al.}(2005)\citenamefont
  {{K\"{o}hl}}, \citenamefont {Moritz}, \citenamefont {{St\"{o}ferle}},
  \citenamefont {{G\"{u}nter}},\ and\ \citenamefont {Esslinger}}]{kohl2005}%
  \BibitemOpen
  \bibfield  {author} {\bibinfo {author} {\bibfnamefont {M.}~\bibnamefont
  {{K\"{o}hl}}}, \bibinfo {author} {\bibfnamefont {H.}~\bibnamefont {Moritz}},
  \bibinfo {author} {\bibfnamefont {T.}~\bibnamefont {{St\"{o}ferle}}},
  \bibinfo {author} {\bibfnamefont {K.}~\bibnamefont {{G\"{u}nter}}}, \ and\
  \bibinfo {author} {\bibfnamefont {T.}~\bibnamefont {Esslinger}},\ }\href
  {http://link.aps.org/doi/10.1103/PhysRevLett.94.080403} {\bibfield  {journal}
  {\bibinfo  {journal} {Phys. Rev. Lett.}\ }\textbf {\bibinfo {volume} {94}},\
  \bibinfo {pages} {080403} (\bibinfo {year} {2005})}\BibitemShut {NoStop}%
\bibitem [{Int()}]{InteractingTI}%
  \BibitemOpen
  \href@noop {} {}\bibinfo {note} {For a review of recent work, see, e.g., M.
  Hohenadler and F. F. Assaad, J. Phys. Condens. Matter {\bf 25}, 143201
  (2013); W. Witczak-Krempa, G. Chen, Y. B. Kim, and L. Balents, Annu. Rev.
  Condens. Matter Phys. {\bf 5}, 57 (2014); T. Senthil, Annu. Rev. Condens.
  Matter Phys. {\bf 6}, 299 (2015); J. Maciejko and G. A. Fiete, Nature Phys.
  {\bf 11}, 385 (2015).}\BibitemShut {Stop}%
\bibitem [{\citenamefont {He}\ \emph {et~al.}(2011{\natexlab{a}})\citenamefont
  {He}, \citenamefont {Kou}, \citenamefont {Liang},\ and\ \citenamefont
  {Feng}}]{he2011}%
  \BibitemOpen
  \bibfield  {author} {\bibinfo {author} {\bibfnamefont {J.}~\bibnamefont
  {He}}, \bibinfo {author} {\bibfnamefont {S.-P.}\ \bibnamefont {Kou}},
  \bibinfo {author} {\bibfnamefont {Y.}~\bibnamefont {Liang}}, \ and\ \bibinfo
  {author} {\bibfnamefont {S.}~\bibnamefont {Feng}},\ }\href
  {http://link.aps.org/doi/10.1103/PhysRevB.83.205116} {\bibfield  {journal}
  {\bibinfo  {journal} {Phys. Rev. B}\ }\textbf {\bibinfo {volume} {83}},\
  \bibinfo {pages} {205116} (\bibinfo {year} {2011}{\natexlab{a}})}\BibitemShut
  {NoStop}%
\bibitem [{\citenamefont {He}\ \emph {et~al.}(2011{\natexlab{b}})\citenamefont
  {He}, \citenamefont {Zong}, \citenamefont {Kou}, \citenamefont {Liang},\ and\
  \citenamefont {Feng}}]{he2011b}%
  \BibitemOpen
  \bibfield  {author} {\bibinfo {author} {\bibfnamefont {J.}~\bibnamefont
  {He}}, \bibinfo {author} {\bibfnamefont {Y.-H.}\ \bibnamefont {Zong}},
  \bibinfo {author} {\bibfnamefont {S.-P.}\ \bibnamefont {Kou}}, \bibinfo
  {author} {\bibfnamefont {Y.}~\bibnamefont {Liang}}, \ and\ \bibinfo {author}
  {\bibfnamefont {S.}~\bibnamefont {Feng}},\ }\href
  {http://link.aps.org/doi/10.1103/PhysRevB.84.035127} {\bibfield  {journal}
  {\bibinfo  {journal} {Phys. Rev. B}\ }\textbf {\bibinfo {volume} {84}},\
  \bibinfo {pages} {035127} (\bibinfo {year} {2011}{\natexlab{b}})}\BibitemShut
  {NoStop}%
\bibitem [{\citenamefont {He}\ \emph {et~al.}(2012{\natexlab{a}})\citenamefont
  {He}, \citenamefont {Liang},\ and\ \citenamefont {Kou}}]{he2012}%
  \BibitemOpen
  \bibfield  {author} {\bibinfo {author} {\bibfnamefont {J.}~\bibnamefont
  {He}}, \bibinfo {author} {\bibfnamefont {Y.}~\bibnamefont {Liang}}, \ and\
  \bibinfo {author} {\bibfnamefont {S.-P.}\ \bibnamefont {Kou}},\ }\href
  {http://link.aps.org/doi/10.1103/PhysRevB.85.205107} {\bibfield  {journal}
  {\bibinfo  {journal} {Phys. Rev. B}\ }\textbf {\bibinfo {volume} {85}},\
  \bibinfo {pages} {205107} (\bibinfo {year} {2012}{\natexlab{a}})}\BibitemShut
  {NoStop}%
\bibitem [{\citenamefont {He}\ \emph {et~al.}(2012{\natexlab{b}})\citenamefont
  {He}, \citenamefont {Wang},\ and\ \citenamefont {Kou}}]{he2012b}%
  \BibitemOpen
  \bibfield  {author} {\bibinfo {author} {\bibfnamefont {J.}~\bibnamefont
  {He}}, \bibinfo {author} {\bibfnamefont {B.}~\bibnamefont {Wang}}, \ and\
  \bibinfo {author} {\bibfnamefont {S.-P.}\ \bibnamefont {Kou}},\ }\href
  {http://link.aps.org/doi/10.1103/PhysRevB.86.235146} {\bibfield  {journal}
  {\bibinfo  {journal} {Phys. Rev. B}\ }\textbf {\bibinfo {volume} {86}},\
  \bibinfo {pages} {235146} (\bibinfo {year} {2012}{\natexlab{b}})}\BibitemShut
  {NoStop}%
\bibitem [{\citenamefont {Maciejko}\ and\ \citenamefont
  {{R\"{u}egg}}(2013)}]{maciejko2013}%
  \BibitemOpen
  \bibfield  {author} {\bibinfo {author} {\bibfnamefont {J.}~\bibnamefont
  {Maciejko}}\ and\ \bibinfo {author} {\bibfnamefont {A.}~\bibnamefont
  {{R\"{u}egg}}},\ }\href {http://link.aps.org/doi/10.1103/PhysRevB.88.241101}
  {\bibfield  {journal} {\bibinfo  {journal} {Phys. Rev. B}\ }\textbf {\bibinfo
  {volume} {88}},\ \bibinfo {pages} {241101} (\bibinfo {year}
  {2013})}\BibitemShut {NoStop}%
\bibitem [{\citenamefont {Zhu}\ \emph {et~al.}(2014)\citenamefont {Zhu},
  \citenamefont {He}, \citenamefont {Zang}, \citenamefont {Liang},\ and\
  \citenamefont {Kou}}]{zhu2014}%
  \BibitemOpen
  \bibfield  {author} {\bibinfo {author} {\bibfnamefont {Y.-X.}\ \bibnamefont
  {Zhu}}, \bibinfo {author} {\bibfnamefont {J.}~\bibnamefont {He}}, \bibinfo
  {author} {\bibfnamefont {C.-L.}\ \bibnamefont {Zang}}, \bibinfo {author}
  {\bibfnamefont {Y.}~\bibnamefont {Liang}}, \ and\ \bibinfo {author}
  {\bibfnamefont {S.-P.}\ \bibnamefont {Kou}},\ }\href
  {http://iopscience.iop.org/article/10.1088/0953-8984/26/17/175601/meta}
  {\bibfield  {journal} {\bibinfo  {journal} {J. Phys. Condens. Matter}\
  }\textbf {\bibinfo {volume} {26}},\ \bibinfo {pages} {175601} (\bibinfo
  {year} {2014})}\BibitemShut {NoStop}%
\bibitem [{\citenamefont {Hickey}\ \emph
  {et~al.}(2015{\natexlab{a}})\citenamefont {Hickey}, \citenamefont {Rath},\
  and\ \citenamefont {Paramekanti}}]{hickey2015}%
  \BibitemOpen
  \bibfield  {author} {\bibinfo {author} {\bibfnamefont {C.}~\bibnamefont
  {Hickey}}, \bibinfo {author} {\bibfnamefont {P.}~\bibnamefont {Rath}}, \ and\
  \bibinfo {author} {\bibfnamefont {A.}~\bibnamefont {Paramekanti}},\ }\href
  {http://link.aps.org/doi/10.1103/PhysRevB.91.134414} {\bibfield  {journal}
  {\bibinfo  {journal} {Phys. Rev. B}\ }\textbf {\bibinfo {volume} {91}},\
  \bibinfo {pages} {134414} (\bibinfo {year} {2015}{\natexlab{a}})}\BibitemShut
  {NoStop}%
\bibitem [{\citenamefont {Zheng}\ \emph {et~al.}(2015)\citenamefont {Zheng},
  \citenamefont {Shen}, \citenamefont {Wang},\ and\ \citenamefont
  {Zhai}}]{zheng2015}%
  \BibitemOpen
  \bibfield  {author} {\bibinfo {author} {\bibfnamefont {W.}~\bibnamefont
  {Zheng}}, \bibinfo {author} {\bibfnamefont {H.}~\bibnamefont {Shen}},
  \bibinfo {author} {\bibfnamefont {Z.}~\bibnamefont {Wang}}, \ and\ \bibinfo
  {author} {\bibfnamefont {H.}~\bibnamefont {Zhai}},\ }\href
  {http://link.aps.org/doi/10.1103/PhysRevB.91.161107} {\bibfield  {journal}
  {\bibinfo  {journal} {Phys. Rev. B}\ }\textbf {\bibinfo {volume} {91}},\
  \bibinfo {pages} {161107} (\bibinfo {year} {2015})}\BibitemShut {NoStop}%
\bibitem [{\citenamefont {Prychynenko}\ and\ \citenamefont
  {Huber}(2016)}]{prychynenko2014}%
  \BibitemOpen
  \bibfield  {author} {\bibinfo {author} {\bibfnamefont {D.}~\bibnamefont
  {Prychynenko}}\ and\ \bibinfo {author} {\bibfnamefont {S.~D.}\ \bibnamefont
  {Huber}},\ }\href {\doibase 10.1016/j.physb.2015.10.027} {\bibfield
  {journal} {\bibinfo  {journal} {Physica B}\ }\textbf {\bibinfo {volume}
  {481}},\ \bibinfo {pages} {53} (\bibinfo {year} {2016})}\BibitemShut
  {NoStop}%
\bibitem [{\citenamefont {Hickey}\ \emph
  {et~al.}(2015{\natexlab{b}})\citenamefont {Hickey}, \citenamefont {Cincio},
  \citenamefont {{Papi\'{c}}},\ and\ \citenamefont
  {Paramekanti}}]{hickey2015b}%
  \BibitemOpen
  \bibfield  {author} {\bibinfo {author} {\bibfnamefont {C.}~\bibnamefont
  {Hickey}}, \bibinfo {author} {\bibfnamefont {L.}~\bibnamefont {Cincio}},
  \bibinfo {author} {\bibfnamefont {Z.}~\bibnamefont {{Papi\'{c}}}}, \ and\
  \bibinfo {author} {\bibfnamefont {A.}~\bibnamefont {Paramekanti}},\ }\href
  {http://arxiv.org/abs/1509.08461} {\bibfield  {journal} {\bibinfo  {journal}
  {arXiv:1509.08461}\ } (\bibinfo {year} {2015}{\natexlab{b}})}\BibitemShut
  {NoStop}%
\bibitem [{\citenamefont {Arun}\ \emph {et~al.}(2015)\citenamefont {Arun},
  \citenamefont {Sohal}, \citenamefont {Hickey},\ and\ \citenamefont
  {Paramekanti}}]{arun2015}%
  \BibitemOpen
  \bibfield  {author} {\bibinfo {author} {\bibfnamefont {V.~S.}\ \bibnamefont
  {Arun}}, \bibinfo {author} {\bibfnamefont {R.}~\bibnamefont {Sohal}},
  \bibinfo {author} {\bibfnamefont {C.}~\bibnamefont {Hickey}}, \ and\ \bibinfo
  {author} {\bibfnamefont {A.}~\bibnamefont {Paramekanti}},\ }\href
  {http://arxiv.org/abs/1510.08856} {\bibfield  {journal} {\bibinfo  {journal}
  {arXiv:1510.08856}\ } (\bibinfo {year} {2015})}\BibitemShut {NoStop}%
\bibitem [{\citenamefont {Hohenadler}\ \emph {et~al.}(2011)\citenamefont
  {Hohenadler}, \citenamefont {Lang},\ and\ \citenamefont
  {Assaad}}]{hohenadler2011}%
  \BibitemOpen
  \bibfield  {author} {\bibinfo {author} {\bibfnamefont {M.}~\bibnamefont
  {Hohenadler}}, \bibinfo {author} {\bibfnamefont {T.~C.}\ \bibnamefont
  {Lang}}, \ and\ \bibinfo {author} {\bibfnamefont {F.~F.}\ \bibnamefont
  {Assaad}},\ }\href {http://link.aps.org/doi/10.1103/PhysRevLett.106.100403}
  {\bibfield  {journal} {\bibinfo  {journal} {Phys. Rev. Lett.}\ }\textbf
  {\bibinfo {volume} {106}},\ \bibinfo {pages} {100403} (\bibinfo {year}
  {2011})}\BibitemShut {NoStop}%
\bibitem [{\citenamefont {Zheng}\ \emph {et~al.}(2011)\citenamefont {Zheng},
  \citenamefont {Zhang},\ and\ \citenamefont {Wu}}]{zheng2011}%
  \BibitemOpen
  \bibfield  {author} {\bibinfo {author} {\bibfnamefont {D.}~\bibnamefont
  {Zheng}}, \bibinfo {author} {\bibfnamefont {G.-M.}\ \bibnamefont {Zhang}}, \
  and\ \bibinfo {author} {\bibfnamefont {C.}~\bibnamefont {Wu}},\ }\href
  {http://link.aps.org/doi/10.1103/PhysRevB.84.205121} {\bibfield  {journal}
  {\bibinfo  {journal} {Phys. Rev. B}\ }\textbf {\bibinfo {volume} {84}},\
  \bibinfo {pages} {205121} (\bibinfo {year} {2011})}\BibitemShut {NoStop}%
\bibitem [{\citenamefont {Hohenadler}\ and\ \citenamefont
  {Assaad}(2012)}]{hohenadler2012}%
  \BibitemOpen
  \bibfield  {author} {\bibinfo {author} {\bibfnamefont {M.}~\bibnamefont
  {Hohenadler}}\ and\ \bibinfo {author} {\bibfnamefont {F.~F.}\ \bibnamefont
  {Assaad}},\ }\href {http://link.aps.org/doi/10.1103/PhysRevB.85.081106}
  {\bibfield  {journal} {\bibinfo  {journal} {Phys. Rev. B}\ }\textbf {\bibinfo
  {volume} {85}},\ \bibinfo {pages} {081106} (\bibinfo {year}
  {2012})}\BibitemShut {NoStop}%
\bibitem [{\citenamefont {Hohenadler}\ \emph {et~al.}(2012)\citenamefont
  {Hohenadler}, \citenamefont {Meng}, \citenamefont {Lang}, \citenamefont
  {Wessel}, \citenamefont {Muramatsu},\ and\ \citenamefont
  {Assaad}}]{hohenadler2012b}%
  \BibitemOpen
  \bibfield  {author} {\bibinfo {author} {\bibfnamefont {M.}~\bibnamefont
  {Hohenadler}}, \bibinfo {author} {\bibfnamefont {Z.~Y.}\ \bibnamefont
  {Meng}}, \bibinfo {author} {\bibfnamefont {T.~C.}\ \bibnamefont {Lang}},
  \bibinfo {author} {\bibfnamefont {S.}~\bibnamefont {Wessel}}, \bibinfo
  {author} {\bibfnamefont {A.}~\bibnamefont {Muramatsu}}, \ and\ \bibinfo
  {author} {\bibfnamefont {F.~F.}\ \bibnamefont {Assaad}},\ }\href
  {http://link.aps.org/doi/10.1103/PhysRevB.85.115132} {\bibfield  {journal}
  {\bibinfo  {journal} {Phys. Rev. B}\ }\textbf {\bibinfo {volume} {85}},\
  \bibinfo {pages} {115132} (\bibinfo {year} {2012})}\BibitemShut {NoStop}%
\bibitem [{\citenamefont {Assaad}\ \emph {et~al.}(2013)\citenamefont {Assaad},
  \citenamefont {Bercx},\ and\ \citenamefont {Hohenadler}}]{assaad2013}%
  \BibitemOpen
  \bibfield  {author} {\bibinfo {author} {\bibfnamefont {F.~F.}\ \bibnamefont
  {Assaad}}, \bibinfo {author} {\bibfnamefont {M.}~\bibnamefont {Bercx}}, \
  and\ \bibinfo {author} {\bibfnamefont {M.}~\bibnamefont {Hohenadler}},\
  }\href {http://link.aps.org/doi/10.1103/PhysRevX.3.011015} {\bibfield
  {journal} {\bibinfo  {journal} {Phys. Rev. X}\ }\textbf {\bibinfo {volume}
  {3}},\ \bibinfo {pages} {011015} (\bibinfo {year} {2013})}\BibitemShut
  {NoStop}%
\bibitem [{\citenamefont {Hung}\ \emph {et~al.}(2013)\citenamefont {Hung},
  \citenamefont {Wang}, \citenamefont {Gu},\ and\ \citenamefont
  {Fiete}}]{hung2013}%
  \BibitemOpen
  \bibfield  {author} {\bibinfo {author} {\bibfnamefont {H.-H.}\ \bibnamefont
  {Hung}}, \bibinfo {author} {\bibfnamefont {L.}~\bibnamefont {Wang}}, \bibinfo
  {author} {\bibfnamefont {Z.-C.}\ \bibnamefont {Gu}}, \ and\ \bibinfo {author}
  {\bibfnamefont {G.~A.}\ \bibnamefont {Fiete}},\ }\href
  {http://link.aps.org/doi/10.1103/PhysRevB.87.121113} {\bibfield  {journal}
  {\bibinfo  {journal} {Phys. Rev. B}\ }\textbf {\bibinfo {volume} {87}},\
  \bibinfo {pages} {121113} (\bibinfo {year} {2013})}\BibitemShut {NoStop}%
\bibitem [{\citenamefont {Lang}\ \emph {et~al.}(2013)\citenamefont {Lang},
  \citenamefont {Essin}, \citenamefont {Gurarie},\ and\ \citenamefont
  {Wessel}}]{lang2013}%
  \BibitemOpen
  \bibfield  {author} {\bibinfo {author} {\bibfnamefont {T.~C.}\ \bibnamefont
  {Lang}}, \bibinfo {author} {\bibfnamefont {A.~M.}\ \bibnamefont {Essin}},
  \bibinfo {author} {\bibfnamefont {V.}~\bibnamefont {Gurarie}}, \ and\
  \bibinfo {author} {\bibfnamefont {S.}~\bibnamefont {Wessel}},\ }\href
  {http://link.aps.org/doi/10.1103/PhysRevB.87.205101} {\bibfield  {journal}
  {\bibinfo  {journal} {Phys. Rev. B}\ }\textbf {\bibinfo {volume} {87}},\
  \bibinfo {pages} {205101} (\bibinfo {year} {2013})}\BibitemShut {NoStop}%
\bibitem [{\citenamefont {Meng}\ \emph {et~al.}(2013)\citenamefont {Meng},
  \citenamefont {Hung},\ and\ \citenamefont {Lang}}]{meng2013}%
  \BibitemOpen
  \bibfield  {author} {\bibinfo {author} {\bibfnamefont {Z.~Y.}\ \bibnamefont
  {Meng}}, \bibinfo {author} {\bibfnamefont {H.-H.}\ \bibnamefont {Hung}}, \
  and\ \bibinfo {author} {\bibfnamefont {T.~C.}\ \bibnamefont {Lang}},\ }\href
  {http://www.worldscientific.com/doi/abs/10.1142/S0217984914300014} {\bibfield
   {journal} {\bibinfo  {journal} {Mod. Phys. Lett. B}\ }\textbf {\bibinfo
  {volume} {28}},\ \bibinfo {pages} {1430001} (\bibinfo {year}
  {2013})}\BibitemShut {NoStop}%
\bibitem [{\citenamefont {Lai}\ and\ \citenamefont {Hung}(2014)}]{lai2014}%
  \BibitemOpen
  \bibfield  {author} {\bibinfo {author} {\bibfnamefont {H.-H.}\ \bibnamefont
  {Lai}}\ and\ \bibinfo {author} {\bibfnamefont {H.-H.}\ \bibnamefont {Hung}},\
  }\href {http://link.aps.org/doi/10.1103/PhysRevB.89.165135} {\bibfield
  {journal} {\bibinfo  {journal} {Phys. Rev. B}\ }\textbf {\bibinfo {volume}
  {89}},\ \bibinfo {pages} {165135} (\bibinfo {year} {2014})}\BibitemShut
  {NoStop}%
\bibitem [{\citenamefont {Hung}\ \emph {et~al.}(2014)\citenamefont {Hung},
  \citenamefont {Chua}, \citenamefont {Wang},\ and\ \citenamefont
  {Fiete}}]{hung2014}%
  \BibitemOpen
  \bibfield  {author} {\bibinfo {author} {\bibfnamefont {H.-H.}\ \bibnamefont
  {Hung}}, \bibinfo {author} {\bibfnamefont {V.}~\bibnamefont {Chua}}, \bibinfo
  {author} {\bibfnamefont {L.}~\bibnamefont {Wang}}, \ and\ \bibinfo {author}
  {\bibfnamefont {G.~A.}\ \bibnamefont {Fiete}},\ }\href
  {http://link.aps.org/doi/10.1103/PhysRevB.89.235104} {\bibfield  {journal}
  {\bibinfo  {journal} {Phys. Rev. B}\ }\textbf {\bibinfo {volume} {89}},\
  \bibinfo {pages} {235104} (\bibinfo {year} {2014})}\BibitemShut {NoStop}%
\bibitem [{\citenamefont {Bercx}\ \emph {et~al.}(2014)\citenamefont {Bercx},
  \citenamefont {Hohenadler},\ and\ \citenamefont {Assaad}}]{bercx2014}%
  \BibitemOpen
  \bibfield  {author} {\bibinfo {author} {\bibfnamefont {M.}~\bibnamefont
  {Bercx}}, \bibinfo {author} {\bibfnamefont {M.}~\bibnamefont {Hohenadler}}, \
  and\ \bibinfo {author} {\bibfnamefont {F.~F.}\ \bibnamefont {Assaad}},\
  }\href {http://link.aps.org/doi/10.1103/PhysRevB.90.075140} {\bibfield
  {journal} {\bibinfo  {journal} {Phys. Rev. B}\ }\textbf {\bibinfo {volume}
  {90}},\ \bibinfo {pages} {075140} (\bibinfo {year} {2014})}\BibitemShut
  {NoStop}%
\bibitem [{\citenamefont {Lai}\ \emph {et~al.}(2014)\citenamefont {Lai},
  \citenamefont {Hung},\ and\ \citenamefont {Fiete}}]{lai2014b}%
  \BibitemOpen
  \bibfield  {author} {\bibinfo {author} {\bibfnamefont {H.-H.}\ \bibnamefont
  {Lai}}, \bibinfo {author} {\bibfnamefont {H.-H.}\ \bibnamefont {Hung}}, \
  and\ \bibinfo {author} {\bibfnamefont {G.~A.}\ \bibnamefont {Fiete}},\ }\href
  {http://link.aps.org/doi/10.1103/PhysRevB.90.195120} {\bibfield  {journal}
  {\bibinfo  {journal} {Phys. Rev. B}\ }\textbf {\bibinfo {volume} {90}},\
  \bibinfo {pages} {195120} (\bibinfo {year} {2014})}\BibitemShut {NoStop}%
\bibitem [{\citenamefont {Ma}\ \emph {et~al.}(2015)\citenamefont {Ma},
  \citenamefont {Lin},\ and\ \citenamefont {Gubernatis}}]{ma2014}%
  \BibitemOpen
  \bibfield  {author} {\bibinfo {author} {\bibfnamefont {T.}~\bibnamefont
  {Ma}}, \bibinfo {author} {\bibfnamefont {H.-Q.}\ \bibnamefont {Lin}}, \ and\
  \bibinfo {author} {\bibfnamefont {J.~E.}\ \bibnamefont {Gubernatis}},\ }\href
  {\doibase 10.1209/0295-5075/111/47003} {\bibfield  {journal} {\bibinfo
  {journal} {Europhys. Lett.}\ }\textbf {\bibinfo {volume} {111}},\ \bibinfo
  {pages} {47003} (\bibinfo {year} {2015})}\BibitemShut {NoStop}%
\bibitem [{\citenamefont {Wu}\ \emph {et~al.}(2015)\citenamefont {Wu},
  \citenamefont {He}, \citenamefont {You}, \citenamefont {Xu}, \citenamefont
  {Meng},\ and\ \citenamefont {Lu}}]{wu2015}%
  \BibitemOpen
  \bibfield  {author} {\bibinfo {author} {\bibfnamefont {H.-Q.}\ \bibnamefont
  {Wu}}, \bibinfo {author} {\bibfnamefont {Y.-Y.}\ \bibnamefont {He}}, \bibinfo
  {author} {\bibfnamefont {Y.-Z.}\ \bibnamefont {You}}, \bibinfo {author}
  {\bibfnamefont {C.}~\bibnamefont {Xu}}, \bibinfo {author} {\bibfnamefont
  {Z.~Y.}\ \bibnamefont {Meng}}, \ and\ \bibinfo {author} {\bibfnamefont
  {Z.-Y.}\ \bibnamefont {Lu}},\ }\href {\doibase 10.1103/PhysRevB.92.165123}
  {\bibfield  {journal} {\bibinfo  {journal} {Phys. Rev. B}\ }\textbf {\bibinfo
  {volume} {92}},\ \bibinfo {pages} {165123} (\bibinfo {year}
  {2015})}\BibitemShut {NoStop}%
\bibitem [{\citenamefont {S\'{e}n\'{e}chal}(2008)}]{senechal2008}%
  \BibitemOpen
  \bibfield  {author} {\bibinfo {author} {\bibfnamefont {D.}~\bibnamefont
  {S\'{e}n\'{e}chal}},\ }\href {http://arxiv.org/abs/0806.2690} {\bibfield
  {journal} {\bibinfo  {journal} {arXiv:0806.2690}\ } (\bibinfo {year}
  {2008})}\BibitemShut {NoStop}%
\bibitem [{\citenamefont {Gros}\ and\ \citenamefont
  {{Valent\'i}}(1993)}]{gros1993}%
  \BibitemOpen
  \bibfield  {author} {\bibinfo {author} {\bibfnamefont {C.}~\bibnamefont
  {Gros}}\ and\ \bibinfo {author} {\bibfnamefont {R.}~\bibnamefont
  {{Valent\'i}}},\ }\href {http://link.aps.org/doi/10.1103/PhysRevB.48.418}
  {\bibfield  {journal} {\bibinfo  {journal} {Phys. Rev. B}\ }\textbf {\bibinfo
  {volume} {48}},\ \bibinfo {pages} {418} (\bibinfo {year} {1993})}\BibitemShut
  {NoStop}%
\bibitem [{\citenamefont {{S\'en\'echal}}\ \emph {et~al.}(2000)\citenamefont
  {{S\'en\'echal}}, \citenamefont {Perez},\ and\ \citenamefont
  {{Pioro-Ladri\`ere}}}]{senechal2000}%
  \BibitemOpen
  \bibfield  {author} {\bibinfo {author} {\bibfnamefont {D.}~\bibnamefont
  {{S\'en\'echal}}}, \bibinfo {author} {\bibfnamefont {D.}~\bibnamefont
  {Perez}}, \ and\ \bibinfo {author} {\bibfnamefont {M.}~\bibnamefont
  {{Pioro-Ladri\`ere}}},\ }\href
  {http://link.aps.org/doi/10.1103/PhysRevLett.84.522} {\bibfield  {journal}
  {\bibinfo  {journal} {Phys. Rev. Lett.}\ }\textbf {\bibinfo {volume} {84}},\
  \bibinfo {pages} {522} (\bibinfo {year} {2000})}\BibitemShut {NoStop}%
\bibitem [{\citenamefont {Potthoff}\ \emph {et~al.}(2003)\citenamefont
  {Potthoff}, \citenamefont {Aichhorn},\ and\ \citenamefont
  {Dahnken}}]{potthoff2003}%
  \BibitemOpen
  \bibfield  {author} {\bibinfo {author} {\bibfnamefont {M.}~\bibnamefont
  {Potthoff}}, \bibinfo {author} {\bibfnamefont {M.}~\bibnamefont {Aichhorn}},
  \ and\ \bibinfo {author} {\bibfnamefont {C.}~\bibnamefont {Dahnken}},\ }\href
  {http://link.aps.org/doi/10.1103/PhysRevLett.91.206402} {\bibfield  {journal}
  {\bibinfo  {journal} {Phys. Rev. Lett.}\ }\textbf {\bibinfo {volume} {91}},\
  \bibinfo {pages} {206402} (\bibinfo {year} {2003})}\BibitemShut {NoStop}%
\bibitem [{\citenamefont {Lichtenstein}\ and\ \citenamefont
  {Katsnelson}(2000)}]{Lichtenstein:2000vn}%
  \BibitemOpen
  \bibfield  {author} {\bibinfo {author} {\bibfnamefont {A.~I.}\ \bibnamefont
  {Lichtenstein}}\ and\ \bibinfo {author} {\bibfnamefont {M.~I.}\ \bibnamefont
  {Katsnelson}},\ }\href {\doibase 10.1103/PhysRevB.62.R9283} {\bibfield
  {journal} {\bibinfo  {journal} {Phys. Rev. B}\ }\textbf {\bibinfo {volume}
  {62}},\ \bibinfo {pages} {R9283} (\bibinfo {year} {2000})}\BibitemShut
  {NoStop}%
\bibitem [{\citenamefont {Kotliar}\ \emph {et~al.}(2001)\citenamefont
  {Kotliar}, \citenamefont {Savrasov}, \citenamefont {{P\'{a}lsson}},\ and\
  \citenamefont {Biroli}}]{Kotliar:2001}%
  \BibitemOpen
  \bibfield  {author} {\bibinfo {author} {\bibfnamefont {G.}~\bibnamefont
  {Kotliar}}, \bibinfo {author} {\bibfnamefont {S.~Y.}\ \bibnamefont
  {Savrasov}}, \bibinfo {author} {\bibfnamefont {G.}~\bibnamefont
  {{P\'{a}lsson}}}, \ and\ \bibinfo {author} {\bibfnamefont {G.}~\bibnamefont
  {Biroli}},\ }\href {\doibase 10.1103/PhysRevLett.87.186401} {\bibfield
  {journal} {\bibinfo  {journal} {Phys. Rev. Lett.}\ }\textbf {\bibinfo
  {volume} {87}},\ \bibinfo {pages} {186401} (\bibinfo {year}
  {2001})}\BibitemShut {NoStop}%
\bibitem [{\citenamefont {Pairault}\ \emph {et~al.}(1998)\citenamefont
  {Pairault}, \citenamefont {{S\'en\'echal}},\ and\ \citenamefont
  {Tremblay}}]{pairault1998}%
  \BibitemOpen
  \bibfield  {author} {\bibinfo {author} {\bibfnamefont {S.}~\bibnamefont
  {Pairault}}, \bibinfo {author} {\bibfnamefont {D.}~\bibnamefont
  {{S\'en\'echal}}}, \ and\ \bibinfo {author} {\bibfnamefont {A.-M.~S.}\
  \bibnamefont {Tremblay}},\ }\href
  {http://link.aps.org/doi/10.1103/PhysRevLett.80.5389} {\bibfield  {journal}
  {\bibinfo  {journal} {Phys. Rev. Lett.}\ }\textbf {\bibinfo {volume} {80}},\
  \bibinfo {pages} {5389} (\bibinfo {year} {1998})}\BibitemShut {NoStop}%
\bibitem [{\citenamefont {Potthoff}(2003)}]{potthoff2003b}%
  \BibitemOpen
  \bibfield  {author} {\bibinfo {author} {\bibfnamefont {M.}~\bibnamefont
  {Potthoff}},\ }\href
  {http://www.springerlink.com/Index/10.1140/epjb/e2003-00121-8} {\bibfield
  {journal} {\bibinfo  {journal} {Eur. Phys. J. B}\ }\textbf {\bibinfo {volume}
  {32}},\ \bibinfo {pages} {429} (\bibinfo {year} {2003})}\BibitemShut
  {NoStop}%
\bibitem [{\citenamefont {Hassan}\ \emph {et~al.}(2013)\citenamefont {Hassan},
  \citenamefont {Goyal}, \citenamefont {Shankar},\ and\ \citenamefont
  {{S\'en\'echal}}}]{hassan2013}%
  \BibitemOpen
  \bibfield  {author} {\bibinfo {author} {\bibfnamefont {S.~R.}\ \bibnamefont
  {Hassan}}, \bibinfo {author} {\bibfnamefont {S.}~\bibnamefont {Goyal}},
  \bibinfo {author} {\bibfnamefont {R.}~\bibnamefont {Shankar}}, \ and\
  \bibinfo {author} {\bibfnamefont {D.}~\bibnamefont {{S\'en\'echal}}},\ }\href
  {http://link.aps.org/doi/10.1103/PhysRevB.88.045301} {\bibfield  {journal}
  {\bibinfo  {journal} {Phys. Rev. B}\ }\textbf {\bibinfo {volume} {88}},\
  \bibinfo {pages} {045301} (\bibinfo {year} {2013})}\BibitemShut {NoStop}%
\bibitem [{\citenamefont {Nguyen}\ and\ \citenamefont
  {Tran}(2013)}]{nguyen2013}%
  \BibitemOpen
  \bibfield  {author} {\bibinfo {author} {\bibfnamefont {H.-S.}\ \bibnamefont
  {Nguyen}}\ and\ \bibinfo {author} {\bibfnamefont {M.-T.}\ \bibnamefont
  {Tran}},\ }\href {http://link.aps.org/doi/10.1103/PhysRevB.88.165132}
  {\bibfield  {journal} {\bibinfo  {journal} {Phys. Rev. B}\ }\textbf {\bibinfo
  {volume} {88}},\ \bibinfo {pages} {165132} (\bibinfo {year}
  {2013})}\BibitemShut {NoStop}%
\bibitem [{\citenamefont {Yu}\ \emph {et~al.}(2011)\citenamefont {Yu},
  \citenamefont {Xie},\ and\ \citenamefont {Li}}]{yu2011}%
  \BibitemOpen
  \bibfield  {author} {\bibinfo {author} {\bibfnamefont {S.-L.}\ \bibnamefont
  {Yu}}, \bibinfo {author} {\bibfnamefont {X.~C.}\ \bibnamefont {Xie}}, \ and\
  \bibinfo {author} {\bibfnamefont {J.-X.}\ \bibnamefont {Li}},\ }\href
  {http://link.aps.org/doi/10.1103/PhysRevLett.107.010401} {\bibfield
  {journal} {\bibinfo  {journal} {Phys. Rev. Lett.}\ }\textbf {\bibinfo
  {volume} {107}},\ \bibinfo {pages} {010401} (\bibinfo {year}
  {2011})}\BibitemShut {NoStop}%
\bibitem [{\citenamefont {Yoshida}\ \emph {et~al.}(2012)\citenamefont
  {Yoshida}, \citenamefont {Fujimoto},\ and\ \citenamefont
  {Kawakami}}]{yoshida2012}%
  \BibitemOpen
  \bibfield  {author} {\bibinfo {author} {\bibfnamefont {T.}~\bibnamefont
  {Yoshida}}, \bibinfo {author} {\bibfnamefont {S.}~\bibnamefont {Fujimoto}}, \
  and\ \bibinfo {author} {\bibfnamefont {N.}~\bibnamefont {Kawakami}},\ }\href
  {http://link.aps.org/doi/10.1103/PhysRevB.85.125113} {\bibfield  {journal}
  {\bibinfo  {journal} {Phys. Rev. B}\ }\textbf {\bibinfo {volume} {85}},\
  \bibinfo {pages} {125113} (\bibinfo {year} {2012})}\BibitemShut {NoStop}%
\bibitem [{\citenamefont {Tada}\ \emph {et~al.}(2012)\citenamefont {Tada},
  \citenamefont {Peters}, \citenamefont {Oshikawa}, \citenamefont {Koga},
  \citenamefont {Kawakami},\ and\ \citenamefont {Fujimoto}}]{tada2012}%
  \BibitemOpen
  \bibfield  {author} {\bibinfo {author} {\bibfnamefont {Y.}~\bibnamefont
  {Tada}}, \bibinfo {author} {\bibfnamefont {R.}~\bibnamefont {Peters}},
  \bibinfo {author} {\bibfnamefont {M.}~\bibnamefont {Oshikawa}}, \bibinfo
  {author} {\bibfnamefont {A.}~\bibnamefont {Koga}}, \bibinfo {author}
  {\bibfnamefont {N.}~\bibnamefont {Kawakami}}, \ and\ \bibinfo {author}
  {\bibfnamefont {S.}~\bibnamefont {Fujimoto}},\ }\href
  {http://link.aps.org/doi/10.1103/PhysRevB.85.165138} {\bibfield  {journal}
  {\bibinfo  {journal} {Phys. Rev. B}\ }\textbf {\bibinfo {volume} {85}},\
  \bibinfo {pages} {165138} (\bibinfo {year} {2012})}\BibitemShut {NoStop}%
\bibitem [{\citenamefont {Cocks}\ \emph {et~al.}(2012)\citenamefont {Cocks},
  \citenamefont {Orth}, \citenamefont {Rachel}, \citenamefont {Buchhold},
  \citenamefont {Le~Hur},\ and\ \citenamefont {Hofstetter}}]{cocks2012}%
  \BibitemOpen
  \bibfield  {author} {\bibinfo {author} {\bibfnamefont {D.}~\bibnamefont
  {Cocks}}, \bibinfo {author} {\bibfnamefont {P.~P.}\ \bibnamefont {Orth}},
  \bibinfo {author} {\bibfnamefont {S.}~\bibnamefont {Rachel}}, \bibinfo
  {author} {\bibfnamefont {M.}~\bibnamefont {Buchhold}}, \bibinfo {author}
  {\bibfnamefont {K.}~\bibnamefont {Le~Hur}}, \ and\ \bibinfo {author}
  {\bibfnamefont {W.}~\bibnamefont {Hofstetter}},\ }\href
  {http://link.aps.org/doi/10.1103/PhysRevLett.109.205303} {\bibfield
  {journal} {\bibinfo  {journal} {Phys. Rev. Lett.}\ }\textbf {\bibinfo
  {volume} {109}},\ \bibinfo {pages} {205303} (\bibinfo {year}
  {2012})}\BibitemShut {NoStop}%
\bibitem [{\citenamefont {Budich}\ \emph {et~al.}(2012)\citenamefont {Budich},
  \citenamefont {Thomale}, \citenamefont {Li}, \citenamefont {Laubach},\ and\
  \citenamefont {Zhang}}]{budich2012}%
  \BibitemOpen
  \bibfield  {author} {\bibinfo {author} {\bibfnamefont {J.~C.}\ \bibnamefont
  {Budich}}, \bibinfo {author} {\bibfnamefont {R.}~\bibnamefont {Thomale}},
  \bibinfo {author} {\bibfnamefont {G.}~\bibnamefont {Li}}, \bibinfo {author}
  {\bibfnamefont {M.}~\bibnamefont {Laubach}}, \ and\ \bibinfo {author}
  {\bibfnamefont {S.-C.}\ \bibnamefont {Zhang}},\ }\href
  {http://link.aps.org/doi/10.1103/PhysRevB.86.201407} {\bibfield  {journal}
  {\bibinfo  {journal} {Phys. Rev. B}\ }\textbf {\bibinfo {volume} {86}},\
  \bibinfo {pages} {201407} (\bibinfo {year} {2012})}\BibitemShut {NoStop}%
\bibitem [{\citenamefont {Wu}\ \emph {et~al.}(2012)\citenamefont {Wu},
  \citenamefont {Rachel}, \citenamefont {Liu},\ and\ \citenamefont
  {Le~Hur}}]{wu2012}%
  \BibitemOpen
  \bibfield  {author} {\bibinfo {author} {\bibfnamefont {W.}~\bibnamefont
  {Wu}}, \bibinfo {author} {\bibfnamefont {S.}~\bibnamefont {Rachel}}, \bibinfo
  {author} {\bibfnamefont {W.-M.}\ \bibnamefont {Liu}}, \ and\ \bibinfo
  {author} {\bibfnamefont {K.}~\bibnamefont {Le~Hur}},\ }\href
  {http://link.aps.org/doi/10.1103/PhysRevB.85.205102} {\bibfield  {journal}
  {\bibinfo  {journal} {Phys. Rev. B}\ }\textbf {\bibinfo {volume} {85}},\
  \bibinfo {pages} {205102} (\bibinfo {year} {2012})}\BibitemShut {NoStop}%
\bibitem [{\citenamefont {Yoshida}\ \emph {et~al.}(2013)\citenamefont
  {Yoshida}, \citenamefont {Peters}, \citenamefont {Fujimoto},\ and\
  \citenamefont {Kawakami}}]{yoshida2013}%
  \BibitemOpen
  \bibfield  {author} {\bibinfo {author} {\bibfnamefont {T.}~\bibnamefont
  {Yoshida}}, \bibinfo {author} {\bibfnamefont {R.}~\bibnamefont {Peters}},
  \bibinfo {author} {\bibfnamefont {S.}~\bibnamefont {Fujimoto}}, \ and\
  \bibinfo {author} {\bibfnamefont {N.}~\bibnamefont {Kawakami}},\ }\href
  {http://link.aps.org/doi/10.1103/PhysRevB.87.085134} {\bibfield  {journal}
  {\bibinfo  {journal} {Phys. Rev. B}\ }\textbf {\bibinfo {volume} {87}},\
  \bibinfo {pages} {085134} (\bibinfo {year} {2013})}\BibitemShut {NoStop}%
\bibitem [{\citenamefont {Miyakoshi}\ and\ \citenamefont
  {Ohta}(2013)}]{miyakoshi2013}%
  \BibitemOpen
  \bibfield  {author} {\bibinfo {author} {\bibfnamefont {S.}~\bibnamefont
  {Miyakoshi}}\ and\ \bibinfo {author} {\bibfnamefont {Y.}~\bibnamefont
  {Ohta}},\ }\href {http://link.aps.org/doi/10.1103/PhysRevB.87.195133}
  {\bibfield  {journal} {\bibinfo  {journal} {Phys. Rev. B}\ }\textbf {\bibinfo
  {volume} {87}},\ \bibinfo {pages} {195133} (\bibinfo {year}
  {2013})}\BibitemShut {NoStop}%
\bibitem [{\citenamefont {Budich}\ \emph {et~al.}(2013)\citenamefont {Budich},
  \citenamefont {Trauzettel},\ and\ \citenamefont {Sangiovanni}}]{budich2013}%
  \BibitemOpen
  \bibfield  {author} {\bibinfo {author} {\bibfnamefont {J.~C.}\ \bibnamefont
  {Budich}}, \bibinfo {author} {\bibfnamefont {B.}~\bibnamefont {Trauzettel}},
  \ and\ \bibinfo {author} {\bibfnamefont {G.}~\bibnamefont {Sangiovanni}},\
  }\href {http://link.aps.org/doi/10.1103/PhysRevB.87.235104} {\bibfield
  {journal} {\bibinfo  {journal} {Phys. Rev. B}\ }\textbf {\bibinfo {volume}
  {87}},\ \bibinfo {pages} {235104} (\bibinfo {year} {2013})}\BibitemShut
  {NoStop}%
\bibitem [{\citenamefont {Nourafkan}\ \emph {et~al.}(2014)\citenamefont
  {Nourafkan}, \citenamefont {Kotliar},\ and\ \citenamefont
  {Tremblay}}]{nourafkan2014}%
  \BibitemOpen
  \bibfield  {author} {\bibinfo {author} {\bibfnamefont {R.}~\bibnamefont
  {Nourafkan}}, \bibinfo {author} {\bibfnamefont {G.}~\bibnamefont {Kotliar}},
  \ and\ \bibinfo {author} {\bibfnamefont {A.-M.~S.}\ \bibnamefont
  {Tremblay}},\ }\href {http://link.aps.org/doi/10.1103/PhysRevB.90.125132}
  {\bibfield  {journal} {\bibinfo  {journal} {Phys. Rev. B}\ }\textbf {\bibinfo
  {volume} {90}},\ \bibinfo {pages} {125132} (\bibinfo {year}
  {2014})}\BibitemShut {NoStop}%
\bibitem [{\citenamefont {Laubach}\ \emph {et~al.}(2014)\citenamefont
  {Laubach}, \citenamefont {Reuther}, \citenamefont {Thomale},\ and\
  \citenamefont {Rachel}}]{laubach2014}%
  \BibitemOpen
  \bibfield  {author} {\bibinfo {author} {\bibfnamefont {M.}~\bibnamefont
  {Laubach}}, \bibinfo {author} {\bibfnamefont {J.}~\bibnamefont {Reuther}},
  \bibinfo {author} {\bibfnamefont {R.}~\bibnamefont {Thomale}}, \ and\
  \bibinfo {author} {\bibfnamefont {S.}~\bibnamefont {Rachel}},\ }\href
  {http://link.aps.org/doi/10.1103/PhysRevB.90.165136} {\bibfield  {journal}
  {\bibinfo  {journal} {Phys. Rev. B}\ }\textbf {\bibinfo {volume} {90}},\
  \bibinfo {pages} {165136} (\bibinfo {year} {2014})}\BibitemShut {NoStop}%
\bibitem [{\citenamefont {Chen}\ \emph {et~al.}(2015)\citenamefont {Chen},
  \citenamefont {Hung}, \citenamefont {Su}, \citenamefont {Fiete},\ and\
  \citenamefont {Ting}}]{chen2015}%
  \BibitemOpen
  \bibfield  {author} {\bibinfo {author} {\bibfnamefont {Y.-H.}\ \bibnamefont
  {Chen}}, \bibinfo {author} {\bibfnamefont {H.-H.}\ \bibnamefont {Hung}},
  \bibinfo {author} {\bibfnamefont {G.}~\bibnamefont {Su}}, \bibinfo {author}
  {\bibfnamefont {G.~A.}\ \bibnamefont {Fiete}}, \ and\ \bibinfo {author}
  {\bibfnamefont {C.~S.}\ \bibnamefont {Ting}},\ }\href
  {http://link.aps.org/doi/10.1103/PhysRevB.91.045122} {\bibfield  {journal}
  {\bibinfo  {journal} {Phys. Rev. B}\ }\textbf {\bibinfo {volume} {91}},\
  \bibinfo {pages} {045122} (\bibinfo {year} {2015})}\BibitemShut {NoStop}%
\bibitem [{\citenamefont {Grandi}\ \emph {et~al.}(2015)\citenamefont {Grandi},
  \citenamefont {Manghi}, \citenamefont {Corradini}, \citenamefont {Bertoni},\
  and\ \citenamefont {Bonini}}]{grandi2015}%
  \BibitemOpen
  \bibfield  {author} {\bibinfo {author} {\bibfnamefont {F.}~\bibnamefont
  {Grandi}}, \bibinfo {author} {\bibfnamefont {F.}~\bibnamefont {Manghi}},
  \bibinfo {author} {\bibfnamefont {O.}~\bibnamefont {Corradini}}, \bibinfo
  {author} {\bibfnamefont {C.~M.}\ \bibnamefont {Bertoni}}, \ and\ \bibinfo
  {author} {\bibfnamefont {A.}~\bibnamefont {Bonini}},\ }\href
  {http://iopscience.iop.org/1367-2630/17/2/023004} {\bibfield  {journal}
  {\bibinfo  {journal} {New J. Phys.}\ }\textbf {\bibinfo {volume} {17}},\
  \bibinfo {pages} {023004} (\bibinfo {year} {2015})}\BibitemShut {NoStop}%
\bibitem [{\citenamefont {Werner}\ and\ \citenamefont
  {Assaad}(2013)}]{werner2013}%
  \BibitemOpen
  \bibfield  {author} {\bibinfo {author} {\bibfnamefont {J.}~\bibnamefont
  {Werner}}\ and\ \bibinfo {author} {\bibfnamefont {F.~F.}\ \bibnamefont
  {Assaad}},\ }\href {http://link.aps.org/doi/10.1103/PhysRevB.88.035113}
  {\bibfield  {journal} {\bibinfo  {journal} {Phys. Rev. B}\ }\textbf {\bibinfo
  {volume} {88}},\ \bibinfo {pages} {035113} (\bibinfo {year}
  {2013})}\BibitemShut {NoStop}%
\bibitem [{\citenamefont {Witczak-Krempa}\ \emph {et~al.}(2014)\citenamefont
  {Witczak-Krempa}, \citenamefont {Knap},\ and\ \citenamefont
  {Abanin}}]{witczak-krempa2014}%
  \BibitemOpen
  \bibfield  {author} {\bibinfo {author} {\bibfnamefont {W.}~\bibnamefont
  {Witczak-Krempa}}, \bibinfo {author} {\bibfnamefont {M.}~\bibnamefont
  {Knap}}, \ and\ \bibinfo {author} {\bibfnamefont {D.}~\bibnamefont
  {Abanin}},\ }\href {http://link.aps.org/doi/10.1103/PhysRevLett.113.136402}
  {\bibfield  {journal} {\bibinfo  {journal} {Phys. Rev. Lett.}\ }\textbf
  {\bibinfo {volume} {113}},\ \bibinfo {pages} {136402} (\bibinfo {year}
  {2014})}\BibitemShut {NoStop}%
\bibitem [{\citenamefont {Kalmeyer}\ and\ \citenamefont
  {Laughlin}(1987)}]{kalmeyer1987}%
  \BibitemOpen
  \bibfield  {author} {\bibinfo {author} {\bibfnamefont {V.}~\bibnamefont
  {Kalmeyer}}\ and\ \bibinfo {author} {\bibfnamefont {R.~B.}\ \bibnamefont
  {Laughlin}},\ }\href {http://link.aps.org/doi/10.1103/PhysRevLett.59.2095}
  {\bibfield  {journal} {\bibinfo  {journal} {Phys. Rev. Lett.}\ }\textbf
  {\bibinfo {volume} {59}},\ \bibinfo {pages} {2095} (\bibinfo {year}
  {1987})}\BibitemShut {NoStop}%
\bibitem [{\citenamefont {Kalmeyer}\ and\ \citenamefont
  {Laughlin}(1989)}]{kalmeyer1989}%
  \BibitemOpen
  \bibfield  {author} {\bibinfo {author} {\bibfnamefont {V.}~\bibnamefont
  {Kalmeyer}}\ and\ \bibinfo {author} {\bibfnamefont {R.~B.}\ \bibnamefont
  {Laughlin}},\ }\href {http://link.aps.org/doi/10.1103/PhysRevB.39.11879}
  {\bibfield  {journal} {\bibinfo  {journal} {Phys. Rev. B}\ }\textbf {\bibinfo
  {volume} {39}},\ \bibinfo {pages} {11879} (\bibinfo {year}
  {1989})}\BibitemShut {NoStop}%
\bibitem [{\citenamefont {Wen}\ \emph {et~al.}(1989)\citenamefont {Wen},
  \citenamefont {Wilczek},\ and\ \citenamefont {Zee}}]{wen1989}%
  \BibitemOpen
  \bibfield  {author} {\bibinfo {author} {\bibfnamefont {X.~G.}\ \bibnamefont
  {Wen}}, \bibinfo {author} {\bibfnamefont {F.}~\bibnamefont {Wilczek}}, \ and\
  \bibinfo {author} {\bibfnamefont {A.}~\bibnamefont {Zee}},\ }\href
  {http://link.aps.org/doi/10.1103/PhysRevB.39.11413} {\bibfield  {journal}
  {\bibinfo  {journal} {Phys. Rev. B}\ }\textbf {\bibinfo {volume} {39}},\
  \bibinfo {pages} {11413} (\bibinfo {year} {1989})}\BibitemShut {NoStop}%
\bibitem [{\citenamefont {Schroeter}\ \emph {et~al.}(2007)\citenamefont
  {Schroeter}, \citenamefont {Kapit}, \citenamefont {Thomale},\ and\
  \citenamefont {Greiter}}]{schroeter2007}%
  \BibitemOpen
  \bibfield  {author} {\bibinfo {author} {\bibfnamefont {D.~F.}\ \bibnamefont
  {Schroeter}}, \bibinfo {author} {\bibfnamefont {E.}~\bibnamefont {Kapit}},
  \bibinfo {author} {\bibfnamefont {R.}~\bibnamefont {Thomale}}, \ and\
  \bibinfo {author} {\bibfnamefont {M.}~\bibnamefont {Greiter}},\ }\href
  {http://link.aps.org/doi/10.1103/PhysRevLett.99.097202} {\bibfield  {journal}
  {\bibinfo  {journal} {Phys. Rev. Lett.}\ }\textbf {\bibinfo {volume} {99}},\
  \bibinfo {pages} {097202} (\bibinfo {year} {2007})}\BibitemShut {NoStop}%
\bibitem [{\citenamefont {Thomale}\ \emph {et~al.}(2009)\citenamefont
  {Thomale}, \citenamefont {Kapit}, \citenamefont {Schroeter},\ and\
  \citenamefont {Greiter}}]{thomale2009}%
  \BibitemOpen
  \bibfield  {author} {\bibinfo {author} {\bibfnamefont {R.}~\bibnamefont
  {Thomale}}, \bibinfo {author} {\bibfnamefont {E.}~\bibnamefont {Kapit}},
  \bibinfo {author} {\bibfnamefont {D.~F.}\ \bibnamefont {Schroeter}}, \ and\
  \bibinfo {author} {\bibfnamefont {M.}~\bibnamefont {Greiter}},\ }\href
  {http://link.aps.org/doi/10.1103/PhysRevB.80.104406} {\bibfield  {journal}
  {\bibinfo  {journal} {Phys. Rev. B}\ }\textbf {\bibinfo {volume} {80}},\
  \bibinfo {pages} {104406} (\bibinfo {year} {2009})}\BibitemShut {NoStop}%
\bibitem [{\citenamefont {Zhang}\ \emph {et~al.}(2011)\citenamefont {Zhang},
  \citenamefont {Grover},\ and\ \citenamefont {Vishwanath}}]{zhang2011}%
  \BibitemOpen
  \bibfield  {author} {\bibinfo {author} {\bibfnamefont {Y.}~\bibnamefont
  {Zhang}}, \bibinfo {author} {\bibfnamefont {T.}~\bibnamefont {Grover}}, \
  and\ \bibinfo {author} {\bibfnamefont {A.}~\bibnamefont {Vishwanath}},\
  }\href {http://link.aps.org/doi/10.1103/PhysRevB.84.075128} {\bibfield
  {journal} {\bibinfo  {journal} {Phys. Rev. B}\ }\textbf {\bibinfo {volume}
  {84}},\ \bibinfo {pages} {075128} (\bibinfo {year} {2011})}\BibitemShut
  {NoStop}%
\bibitem [{\citenamefont {Nielsen}\ \emph {et~al.}(2013)\citenamefont
  {Nielsen}, \citenamefont {Sierra},\ and\ \citenamefont
  {Cirac}}]{nielsen2013}%
  \BibitemOpen
  \bibfield  {author} {\bibinfo {author} {\bibfnamefont {A.~E.~B.}\
  \bibnamefont {Nielsen}}, \bibinfo {author} {\bibfnamefont {G.}~\bibnamefont
  {Sierra}}, \ and\ \bibinfo {author} {\bibfnamefont {J.~I.}\ \bibnamefont
  {Cirac}},\ }\href
  {http://www.nature.com/ncomms/2013/131128/ncomms3864/full/ncomms3864.html}
  {\bibfield  {journal} {\bibinfo  {journal} {Nature Commun.}\ }\textbf
  {\bibinfo {volume} {4}},\ \bibinfo {pages} {2864} (\bibinfo {year}
  {2013})}\BibitemShut {NoStop}%
\bibitem [{\citenamefont {Sorella}\ \emph {et~al.}(2012)\citenamefont
  {Sorella}, \citenamefont {Otsuka},\ and\ \citenamefont
  {Yunoki}}]{sorella2012}%
  \BibitemOpen
  \bibfield  {author} {\bibinfo {author} {\bibfnamefont {S.}~\bibnamefont
  {Sorella}}, \bibinfo {author} {\bibfnamefont {Y.}~\bibnamefont {Otsuka}}, \
  and\ \bibinfo {author} {\bibfnamefont {S.}~\bibnamefont {Yunoki}},\ }\href
  {http://www.nature.com/srep/2012/121218/srep00992/full/srep00992.html}
  {\bibfield  {journal} {\bibinfo  {journal} {Sci. Rep.}\ }\textbf {\bibinfo
  {volume} {2}},\ \bibinfo {pages} {992} (\bibinfo {year} {2012})}\BibitemShut
  {NoStop}%
\bibitem [{\citenamefont {Luttinger}\ and\ \citenamefont
  {Ward}(1960)}]{luttinger1960}%
  \BibitemOpen
  \bibfield  {author} {\bibinfo {author} {\bibfnamefont {J.~M.}\ \bibnamefont
  {Luttinger}}\ and\ \bibinfo {author} {\bibfnamefont {J.~C.}\ \bibnamefont
  {Ward}},\ }\href {http://link.aps.org/doi/10.1103/PhysRev.118.1417}
  {\bibfield  {journal} {\bibinfo  {journal} {Phys. Rev.}\ }\textbf {\bibinfo
  {volume} {118}},\ \bibinfo {pages} {1417} (\bibinfo {year}
  {1960})}\BibitemShut {NoStop}%
\bibitem [{\citenamefont {Aichhorn}\ \emph {et~al.}(2006)\citenamefont
  {Aichhorn}, \citenamefont {Arrigoni}, \citenamefont {Potthoff},\ and\
  \citenamefont {Hanke}}]{aichhorn2006}%
  \BibitemOpen
  \bibfield  {author} {\bibinfo {author} {\bibfnamefont {M.}~\bibnamefont
  {Aichhorn}}, \bibinfo {author} {\bibfnamefont {E.}~\bibnamefont {Arrigoni}},
  \bibinfo {author} {\bibfnamefont {M.}~\bibnamefont {Potthoff}}, \ and\
  \bibinfo {author} {\bibfnamefont {W.}~\bibnamefont {Hanke}},\ }\href
  {http://link.aps.org/doi/10.1103/PhysRevB.74.024508} {\bibfield  {journal}
  {\bibinfo  {journal} {Phys. Rev. B}\ }\textbf {\bibinfo {volume} {74}},\
  \bibinfo {pages} {024508} (\bibinfo {year} {2006})}\BibitemShut {NoStop}%
\bibitem [{\citenamefont {Semenoff}(1984)}]{semenoff1984}%
  \BibitemOpen
  \bibfield  {author} {\bibinfo {author} {\bibfnamefont {G.~W.}\ \bibnamefont
  {Semenoff}},\ }\href {http://link.aps.org/doi/10.1103/PhysRevLett.53.2449}
  {\bibfield  {journal} {\bibinfo  {journal} {Phys. Rev. Lett.}\ }\textbf
  {\bibinfo {volume} {53}},\ \bibinfo {pages} {2449} (\bibinfo {year}
  {1984})}\BibitemShut {NoStop}%
\bibitem [{\citenamefont {Niu}\ \emph {et~al.}(1985)\citenamefont {Niu},
  \citenamefont {Thouless},\ and\ \citenamefont {Wu}}]{niu1985}%
  \BibitemOpen
  \bibfield  {author} {\bibinfo {author} {\bibfnamefont {Q.}~\bibnamefont
  {Niu}}, \bibinfo {author} {\bibfnamefont {D.~J.}\ \bibnamefont {Thouless}}, \
  and\ \bibinfo {author} {\bibfnamefont {Y.-S.}\ \bibnamefont {Wu}},\ }\href
  {http://link.aps.org/doi/10.1103/PhysRevB.31.3372} {\bibfield  {journal}
  {\bibinfo  {journal} {Phys. Rev. B}\ }\textbf {\bibinfo {volume} {31}},\
  \bibinfo {pages} {3372} (\bibinfo {year} {1985})}\BibitemShut {NoStop}%
\bibitem [{\citenamefont {Volovik}(2003)}]{VolovikBook}%
  \BibitemOpen
  \bibfield  {author} {\bibinfo {author} {\bibfnamefont {G.~E.}\ \bibnamefont
  {Volovik}},\ }\href@noop {} {\emph {\bibinfo {title} {The Universe in a
  Helium Droplet}}}\ (\bibinfo  {publisher} {Oxford University Press},\
  \bibinfo {address} {Oxford},\ \bibinfo {year} {2003})\BibitemShut {NoStop}%
\bibitem [{\citenamefont {Wang}\ \emph {et~al.}(2010)\citenamefont {Wang},
  \citenamefont {Qi},\ and\ \citenamefont {Zhang}}]{wang2010}%
  \BibitemOpen
  \bibfield  {author} {\bibinfo {author} {\bibfnamefont {Z.}~\bibnamefont
  {Wang}}, \bibinfo {author} {\bibfnamefont {X.-L.}\ \bibnamefont {Qi}}, \ and\
  \bibinfo {author} {\bibfnamefont {S.-C.}\ \bibnamefont {Zhang}},\ }\href
  {http://link.aps.org/doi/10.1103/PhysRevLett.105.256803} {\bibfield
  {journal} {\bibinfo  {journal} {Phys. Rev. Lett.}\ }\textbf {\bibinfo
  {volume} {105}},\ \bibinfo {pages} {256803} (\bibinfo {year}
  {2010})}\BibitemShut {NoStop}%
\bibitem [{\citenamefont {Wang}\ and\ \citenamefont {Zhang}(2012)}]{wang2012}%
  \BibitemOpen
  \bibfield  {author} {\bibinfo {author} {\bibfnamefont {Z.}~\bibnamefont
  {Wang}}\ and\ \bibinfo {author} {\bibfnamefont {S.-C.}\ \bibnamefont
  {Zhang}},\ }\href {http://link.aps.org/doi/10.1103/PhysRevX.2.031008}
  {\bibfield  {journal} {\bibinfo  {journal} {Phys. Rev. X}\ }\textbf {\bibinfo
  {volume} {2}},\ \bibinfo {pages} {031008} (\bibinfo {year}
  {2012})}\BibitemShut {NoStop}%
\bibitem [{\citenamefont {Fukui}\ \emph {et~al.}(2005)\citenamefont {Fukui},
  \citenamefont {Hatsugai},\ and\ \citenamefont {Suzuki}}]{Fukui:2005fk}%
  \BibitemOpen
  \bibfield  {author} {\bibinfo {author} {\bibfnamefont {T.}~\bibnamefont
  {Fukui}}, \bibinfo {author} {\bibfnamefont {Y.}~\bibnamefont {Hatsugai}}, \
  and\ \bibinfo {author} {\bibfnamefont {H.}~\bibnamefont {Suzuki}},\ }\href
  {http://jpsj.ipap.jp/link?JPSJ/74/1674/} {\bibfield  {journal} {\bibinfo
  {journal} {J. Phys. Soc. Jpn.}\ }\textbf {\bibinfo {volume} {74}},\ \bibinfo
  {pages} {1674} (\bibinfo {year} {2005})}\BibitemShut {NoStop}%
\bibitem [{\citenamefont {Thouless}\ \emph {et~al.}(1982)\citenamefont
  {Thouless}, \citenamefont {Kohmoto}, \citenamefont {Nightingale},\ and\
  \citenamefont {den Nijs}}]{thouless1982}%
  \BibitemOpen
  \bibfield  {author} {\bibinfo {author} {\bibfnamefont {D.~J.}\ \bibnamefont
  {Thouless}}, \bibinfo {author} {\bibfnamefont {M.}~\bibnamefont {Kohmoto}},
  \bibinfo {author} {\bibfnamefont {M.~P.}\ \bibnamefont {Nightingale}}, \ and\
  \bibinfo {author} {\bibfnamefont {M.}~\bibnamefont {den Nijs}},\ }\href
  {http://link.aps.org/doi/10.1103/PhysRevLett.49.405} {\bibfield  {journal}
  {\bibinfo  {journal} {Phys. Rev. Lett.}\ }\textbf {\bibinfo {volume} {49}},\
  \bibinfo {pages} {405} (\bibinfo {year} {1982})}\BibitemShut {NoStop}%
\bibitem [{\citenamefont {Gurarie}\ and\ \citenamefont
  {Essin}(2013)}]{gurarie2013}%
  \BibitemOpen
  \bibfield  {author} {\bibinfo {author} {\bibfnamefont {V.}~\bibnamefont
  {Gurarie}}\ and\ \bibinfo {author} {\bibfnamefont {A.~M.}\ \bibnamefont
  {Essin}},\ }\href
  {http://link.springer.com/article/10.1134/S0021364013040061} {\bibfield
  {journal} {\bibinfo  {journal} {JETP Lett.}\ }\textbf {\bibinfo {volume}
  {97}},\ \bibinfo {pages} {233} (\bibinfo {year} {2013})}\BibitemShut
  {NoStop}%
\bibitem [{\citenamefont {Gurarie}(2011)}]{gurarie2011}%
  \BibitemOpen
  \bibfield  {author} {\bibinfo {author} {\bibfnamefont {V.}~\bibnamefont
  {Gurarie}},\ }\href {http://link.aps.org/doi/10.1103/PhysRevB.83.085426}
  {\bibfield  {journal} {\bibinfo  {journal} {Phys. Rev. B}\ }\textbf {\bibinfo
  {volume} {83}},\ \bibinfo {pages} {085426} (\bibinfo {year}
  {2011})}\BibitemShut {NoStop}%
\bibitem [{\citenamefont {Bolech}\ \emph {et~al.}(2003)\citenamefont {Bolech},
  \citenamefont {Kancharla},\ and\ \citenamefont {Kotliar}}]{Bolech:2003ye}%
  \BibitemOpen
  \bibfield  {author} {\bibinfo {author} {\bibfnamefont {C.~J.}\ \bibnamefont
  {Bolech}}, \bibinfo {author} {\bibfnamefont {S.~S.}\ \bibnamefont
  {Kancharla}}, \ and\ \bibinfo {author} {\bibfnamefont {G.}~\bibnamefont
  {Kotliar}},\ }\href {\doibase 10.1103/PhysRevB.67.075110} {\bibfield
  {journal} {\bibinfo  {journal} {Phys. Rev. B}\ }\textbf {\bibinfo {volume}
  {67}},\ \bibinfo {pages} {075110} (\bibinfo {year} {2003})}\BibitemShut
  {NoStop}%
\bibitem [{\citenamefont {Kancharla}\ \emph {et~al.}(2008)\citenamefont
  {Kancharla}, \citenamefont {Kyung}, \citenamefont {S{\'e}n{\'e}chal},
  \citenamefont {Civelli}, \citenamefont {Capone}, \citenamefont {Kotliar},\
  and\ \citenamefont {Tremblay}}]{Kancharla:2008vn}%
  \BibitemOpen
  \bibfield  {author} {\bibinfo {author} {\bibfnamefont {S.~S.}\ \bibnamefont
  {Kancharla}}, \bibinfo {author} {\bibfnamefont {B.}~\bibnamefont {Kyung}},
  \bibinfo {author} {\bibfnamefont {D.}~\bibnamefont {S{\'e}n{\'e}chal}},
  \bibinfo {author} {\bibfnamefont {M.}~\bibnamefont {Civelli}}, \bibinfo
  {author} {\bibfnamefont {M.}~\bibnamefont {Capone}}, \bibinfo {author}
  {\bibfnamefont {G.}~\bibnamefont {Kotliar}}, \ and\ \bibinfo {author}
  {\bibfnamefont {A.-M.~S.}\ \bibnamefont {Tremblay}},\ }\href {\doibase
  10.1103/PhysRevB.77.184516} {\bibfield  {journal} {\bibinfo  {journal} {Phys.
  Rev. B}\ }\textbf {\bibinfo {volume} {77}},\ \bibinfo {pages} {184516}
  (\bibinfo {year} {2008})}\BibitemShut {NoStop}%
\bibitem [{\citenamefont {S{\'e}n{\'e}chal}(2015)}]{Senechal:2015rm}%
  \BibitemOpen
  \bibfield  {author} {\bibinfo {author} {\bibfnamefont {D.}~\bibnamefont
  {S{\'e}n{\'e}chal}},\ }\enquote {\bibinfo {title} {Quantum cluster methods:
  {CPT} and {CDMFT}},}\ in\ \href@noop {} {\emph {\bibinfo {booktitle}
  {Many-Body Physics: From Kondo to Hubbard, Lecture Notes of the Autumn School
  on Correlated Electrons 2015}}},\ Vol.~\bibinfo {volume} {5},\ \bibinfo
  {editor} {edited by\ \bibinfo {editor} {\bibfnamefont {E.}~\bibnamefont
  {Pavarini}}, \bibinfo {editor} {\bibfnamefont {E.}~\bibnamefont {Koch}}, \
  and\ \bibinfo {editor} {\bibfnamefont {P.}~\bibnamefont {Coleman}}}\
  (\bibinfo  {publisher} {Forschungszentrum J{\"u}lich},\ \bibinfo {year}
  {2015})\ Chap.~\bibinfo {chapter} {13}, pp.\ \bibinfo {pages} {13.1 --
  13.32}\BibitemShut {NoStop}%
\bibitem [{\citenamefont {Zhang}\ \emph {et~al.}(2012)\citenamefont {Zhang},
  \citenamefont {Grover}, \citenamefont {Turner}, \citenamefont {Oshikawa},\
  and\ \citenamefont {Vishwanath}}]{zhang2012}%
  \BibitemOpen
  \bibfield  {author} {\bibinfo {author} {\bibfnamefont {Y.}~\bibnamefont
  {Zhang}}, \bibinfo {author} {\bibfnamefont {T.}~\bibnamefont {Grover}},
  \bibinfo {author} {\bibfnamefont {A.}~\bibnamefont {Turner}}, \bibinfo
  {author} {\bibfnamefont {M.}~\bibnamefont {Oshikawa}}, \ and\ \bibinfo
  {author} {\bibfnamefont {A.}~\bibnamefont {Vishwanath}},\ }\href {\doibase
  10.1103/PhysRevB.85.235151} {\bibfield  {journal} {\bibinfo  {journal} {Phys.
  Rev. B}\ }\textbf {\bibinfo {volume} {85}},\ \bibinfo {pages} {235151}
  (\bibinfo {year} {2012})}\BibitemShut {NoStop}%
\bibitem [{\citenamefont {Motrunich}(2005)}]{motrunich2005}%
  \BibitemOpen
  \bibfield  {author} {\bibinfo {author} {\bibfnamefont {O.~I.}\ \bibnamefont
  {Motrunich}},\ }\href {http://link.aps.org/doi/10.1103/PhysRevB.72.045105}
  {\bibfield  {journal} {\bibinfo  {journal} {Phys. Rev. B}\ }\textbf {\bibinfo
  {volume} {72}},\ \bibinfo {pages} {045105} (\bibinfo {year}
  {2005})}\BibitemShut {NoStop}%
\bibitem [{\citenamefont {Lee}\ and\ \citenamefont {Lee}(2005)}]{lee2005}%
  \BibitemOpen
  \bibfield  {author} {\bibinfo {author} {\bibfnamefont {S.-S.}\ \bibnamefont
  {Lee}}\ and\ \bibinfo {author} {\bibfnamefont {P.~A.}\ \bibnamefont {Lee}},\
  }\href {http://link.aps.org/doi/10.1103/PhysRevLett.95.036403} {\bibfield
  {journal} {\bibinfo  {journal} {Phys. Rev. Lett.}\ }\textbf {\bibinfo
  {volume} {95}},\ \bibinfo {pages} {036403} (\bibinfo {year}
  {2005})}\BibitemShut {NoStop}%
\bibitem [{\citenamefont {Raghu}\ \emph {et~al.}(2008)\citenamefont {Raghu},
  \citenamefont {Qi}, \citenamefont {Honerkamp},\ and\ \citenamefont
  {Zhang}}]{raghu2008}%
  \BibitemOpen
  \bibfield  {author} {\bibinfo {author} {\bibfnamefont {S.}~\bibnamefont
  {Raghu}}, \bibinfo {author} {\bibfnamefont {X.-L.}\ \bibnamefont {Qi}},
  \bibinfo {author} {\bibfnamefont {C.}~\bibnamefont {Honerkamp}}, \ and\
  \bibinfo {author} {\bibfnamefont {S.-C.}\ \bibnamefont {Zhang}},\ }\href
  {http://link.aps.org/doi/10.1103/PhysRevLett.100.156401} {\bibfield
  {journal} {\bibinfo  {journal} {Phys. Rev. Lett.}\ }\textbf {\bibinfo
  {volume} {100}},\ \bibinfo {pages} {156401} (\bibinfo {year}
  {2008})}\BibitemShut {NoStop}%
\bibitem [{\citenamefont {Faye}\ \emph {et~al.}(2014)\citenamefont {Faye},
  \citenamefont {{S\'en\'echal}},\ and\ \citenamefont {Hassan}}]{faye2014}%
  \BibitemOpen
  \bibfield  {author} {\bibinfo {author} {\bibfnamefont {J.~P.~L.}\
  \bibnamefont {Faye}}, \bibinfo {author} {\bibfnamefont {D.}~\bibnamefont
  {{S\'en\'echal}}}, \ and\ \bibinfo {author} {\bibfnamefont {S.~R.}\
  \bibnamefont {Hassan}},\ }\href
  {http://link.aps.org/doi/10.1103/PhysRevB.89.115130} {\bibfield  {journal}
  {\bibinfo  {journal} {Phys. Rev. B}\ }\textbf {\bibinfo {volume} {89}},\
  \bibinfo {pages} {115130} (\bibinfo {year} {2014})}\BibitemShut {NoStop}%
\bibitem [{\citenamefont {Alba}\ \emph {et~al.}(2015)\citenamefont {Alba},
  \citenamefont {Pachos},\ and\ \citenamefont {Garcia-Ripoll}}]{alba2015}%
  \BibitemOpen
  \bibfield  {author} {\bibinfo {author} {\bibfnamefont {E.}~\bibnamefont
  {Alba}}, \bibinfo {author} {\bibfnamefont {J.}~\bibnamefont {Pachos}}, \ and\
  \bibinfo {author} {\bibfnamefont {J.~J.}\ \bibnamefont {Garcia-Ripoll}},\
  }\href {http://arxiv.org/abs/1505.00374} {\bibfield  {journal} {\bibinfo
  {journal} {arXiv:1505.00374}\ } (\bibinfo {year} {2015})}\BibitemShut
  {NoStop}%
\bibitem [{\citenamefont {Grushin}\ \emph {et~al.}(2015)\citenamefont
  {Grushin}, \citenamefont {Motruk}, \citenamefont {Zaletel},\ and\
  \citenamefont {Pollmann}}]{grushin2015}%
  \BibitemOpen
  \bibfield  {author} {\bibinfo {author} {\bibfnamefont {A.~G.}\ \bibnamefont
  {Grushin}}, \bibinfo {author} {\bibfnamefont {J.}~\bibnamefont {Motruk}},
  \bibinfo {author} {\bibfnamefont {M.~P.}\ \bibnamefont {Zaletel}}, \ and\
  \bibinfo {author} {\bibfnamefont {F.}~\bibnamefont {Pollmann}},\ }\href
  {http://link.aps.org/doi/10.1103/PhysRevB.91.035136} {\bibfield  {journal}
  {\bibinfo  {journal} {Phys. Rev. B}\ }\textbf {\bibinfo {volume} {91}},\
  \bibinfo {pages} {035136} (\bibinfo {year} {2015})}\BibitemShut {NoStop}%
\end{thebibliography}%

\end{document}